\documentclass[aps,pra,twocolumn,amsmath,amssymb,nofootinbib,showpacs,superscriptaddress]{revtex4-2}
\usepackage[english]{babel}
\usepackage{latexsym}
\usepackage{graphics}
\usepackage{graphicx}
\usepackage{epsfig}
\usepackage{color}
\usepackage{bm}
\usepackage{amsmath}
\usepackage{amssymb}
\usepackage{amsthm}
\usepackage{dcolumn}
\usepackage{bm}
\usepackage{float}
\usepackage{hyperref}
\usepackage{color}
\usepackage{epstopdf}
\usepackage{braket}
\usepackage{cleveref}
\usepackage[svgnames]{xcolor}
\hypersetup{hidelinks,colorlinks=true,allcolors=DarkBlue}

\theoremstyle{remark}

\begin{document}
	
	\preprint{APS/123-QED}
	
	\title{Realizing a class of stabilizer quantum error correction codes \\ using a single ancilla and circular connectivity}
	
	\author{A.V. Antipov}
	\email{aantipov@nes.ru}
	\affiliation{Russian Quantum Center, Skolkovo, Moscow 143025, Russia}
	\affiliation{National University of Science and Technology ``MISIS'', 119049 Moscow, Russia}
	\author{E.O. Kiktenko}
	\affiliation{Russian Quantum Center, Skolkovo, Moscow 143025, Russia}
	\affiliation{National University of Science and Technology ``MISIS'', 119049 Moscow, Russia}
	\author{A.K. Fedorov}
	\affiliation{Russian Quantum Center, Skolkovo, Moscow 143025, Russia}
	\affiliation{National University of Science and Technology ``MISIS'', 119049 Moscow, Russia}

	\date{\today}
	\begin{abstract}
		We describe a class of ``neighboring-blocks`` stabilizer quantum error correction codes and demonstrate that such class of codes can be implemented in 
		a resource-efficient manner using a single ancilla and circular near-neighbor qubit connectivity. 
		We propose an implementation for syndrome-measurement circuits for codes from the class and illustrate its workings for cases of {3-qubit repetition code, Laflamme's 5-qubit code, and Shor's 9-qubit code.} 
		For {3-qubit repetition code and Laflamme's 5-qubit code} suggested scheme has the property that it uses only native two-qubit CNS (\textsf{CNOT}-\textsf{SWAP}) gates, 
		which potentially reduces the amount of non-correctable errors due to the shorter gate time. 
		{Elements of the scheme can be used to implement surface code with near-neighbour connectivity using single ancilla, as demonstrated in an example.}
		We developed efficient decoding procedures for repetition codes and the {Laflamme's 5-qubit code} using a minimum weight-perfect matching approach to account for the specific order of measurements in our scheme.
		The analysis of noise levels for which the scheme could show improvements in the fidelity of a stored logical {qubit} in the {3-qubit repetition code} and {Laflamme's 5-qubit code} cases is provided. 
		We complement our results by realizing the developed scheme for a 3-qubit code using an IBM quantum processor and the {Laflamme's 5-qubit code} using the state-vector simulator. 
	\end{abstract}
	%and CNS two-qubit gates which are native for superconducting architectures. 
	\maketitle

	\section{Introduction}
	
	Quantum computers have the potential to outperform computing devices based on classical principles in a number of problems ranging from cryptoanalysis to chemistry simulation. 
	During the last decades, key building blocks of quantum processors and simulators based on various physical principles have been demonstrated~\cite{Lukin2017,Monroe2017,Blatt2018-2,Browaeys2018,Martinis2019,Monroe2019,Browaeys2020,Blatt2021-3}. 
	However, exploring the full potential of quantum computing requires reducing uncontrollable environment effects (decoherence) that cause errors.
	At the early stage of quantum computing, the accumulation of error caused by decoherence was used to argue that it is infeasible to build a large-scale quantum computer~\cite{Unruh1995,Landauer1995}.
	Remarkably, a possible solution for overcoming this challenge, which is to use quantum error correction, has been suggested (for a review, see Refs.~\cite{Roffe2019,Georgescu2020}).
	The development of quantum error correction codes is complicated due to quantum no-cloning theorem. 
	Still, the idea behind error correction in the quantum domain is to use many (imperfect) physical qubits to encode one (perfect) logical qubit and perform measurements in order to detect and correct errors. 
	This framework has been actively used to develop various types of quantum error correction techniques~\cite{Shor1995,Shor1996,Steane1996,Gottesman1997,Aharonov2008,Fowler2009,Fowler2011,Fowler2012}. 
	Basic physical principles of quantum error correction have been also demonstrated~\cite{Wineland2004,Blatt2011,Martinis2014}, 
	and existing noisy intermediate-scale quantum devices have become 
	a full-fledged testbed for experimental demonstration of error-correction codes~\cite{Wallraf2019,Wallraf2020,Blatt2020,Blatt2021-2,Monroe2021,Martinis2021,Kelly2021,Lukin2022,Pan2022}.
	However, there are challenges in demonstrating even smallest and simplest prototypes of error-correction codes, since existing quantum error correction tools in their straightforward realizations require resources, 
	such as significantly high number of required qubits and the need of a complex connectivity patterns between qubits.
	
	The number of required qubits is high not only because of the required overhead in the number of physical data qubits for encoding logical {qubits}, 
	but because of additional qubits, known as ancillas, that are used for extracting syndrome measurements that will be used in the process of correcting errors. 
	Connections between qubits, in turn, should allow executing algorithms for performing syndrome measurements. 
	However, even the smallest error correction codes require complex connectivity patters. 
	For a number of quantum computing platforms, specifically, superconducting qubits, ensuring required topology of connections may be a serious engineering challenge~\cite{Martinis2004,Oliver2020}.
	That is why many superconducting quantum computing devices at hand do not have enough resources for demonstrating even simple prototypes of codes that, for instance, correct arbitrary single-qubit errors. 
	Thus, an ability to use as few connections and qubits as possible in syndrome measurement circuits is important for a proof-of-principle implementations of error correction codes.
	
	In the present work, we address the problem of reducing resources for implementing quantum error-correcting codes for a particular class of stabilizer codes.  
	As we find, there is a class of stabilizer codes, which we refer to as ``codes with neigboring blocks``, that can be realized using single ancilla and circular pattern of connectivity.
	Moreover, our scheme for their implementation can use native CNS two-qubit gates.
	The number of qubits for implementing codes from the class is irreducible, since there is no way to use less than one ancilla qubit for syndrome measurements and the number of qubits for encoding data is fixed by a code's mathematical structure. 
	The circular pattern of connectivity requires only two physical connections for every qubit in a system. 
	We expect that such properties can be of interest for particular classes of quantum computing devices, such as superconducting circuits, and that is why we use native two-qubit gates for this platform in our consideration. 
	Native CNS gates are less noisy in superconducting architecture~\cite{Siewert2003}, and thus using them instead of CNOT gates in the implementation of error correction code is potentially advantageous.
	
	Notable examples of stabilizer codes from the class under consideration are repetition codes, {the Laflamme's 5-qubit code~\cite{https://doi.org/10.48550/arxiv.quant-ph/9602019}, and the Shor's 9-qubit code~\cite{Shor1995}}. 
	Repetition codes correct bit-flip errors, whereas five- and nine-qubit codes correct arbitrary single-qubit errors, and thus these codes represent a significant deal of interest in the context of experimental realizations.
	Here we suggest a general scheme for implementing circuits for syndrome measurements in the ``neighboring blocks'' class of codes and illustrate its workings using simulations 
	(for the 3-qubit repetition code and the Laflamme's 5-qubit code) and experiments on the IBM quantum processor (for the 3-qubit repetition code using only four physical qubits).
	
	This paper is organized as follows. 
	In Sec.~\ref{sec:stab-codes}, we provide a general background including basic introduction to stabilizer codes and description of the CNS gate. 
	In Sec.~\ref{sec:single-ancilla}, we describe the suggested approach and illustrate it for cases of 3-qubit repetition code, Laflamme's 5-qubit code, and Shor's 9-qubit code. 
	In Sec.~\ref{sec:generalization}, we provide a generalization of the scheme for the $n$-qubit case. 
	Sec.~\ref{sec:decoding} contains a description of the efficient procedures for decoding measurement syndromes obtained from repetition codes and the Laflamme's 5-qubit code for the proposed scheme. 
	Our analysis of error levels for which the 3-qubit repetition code and the Laflamme's 5-qubit code have improvements for storing logical qubit is provided in Sec.~\ref{sec:simulations}. 
	Sec.~\ref{sec:IBM} contains description and results of experiments using IBM quantum processor for the realization of the 3-qubit repetition code using our scheme. 
	We summarize our results and conclude in Sec.~\ref{sec:conlcusion}.
	
	\section{Background}\label{sec:stab-codes}
	
	Here we briefly outline features of stabilizer codes. 
	A more general explanation can be found in Refs.~\cite{Gottesman1997,Roffe2019}. 
	Additionally, we clarify how exactly measurements defined by an operator from {generating set} are performed using quantum circuits.
	We also shortly discuss the implementation of required quantum circuits using the CNS gate, which is of interest for superconducting quantum computing platform. 
	
	\subsection{Stabilizer codes}
	
	To specify a quantum error correction code, one should specify a code space $C$. 
	A code space is a subspace of the Hilbert space constituting all possible quantum states of the composite system. 
	For instance, if we want to encode $k$ logical qubits into $n$ physical qubits, then the dimensionality of the Hilbert space for the composite system would be $2^n$, and the dimensionality of the code space would be $2^k$, where $k < n$. 
	In the case of stabilizer codes, the corresponding code space is defined by the stabilizer $S$. 
	Before going into details in defining a code space $C$ by a stabilizer $S$, let us define stabilizer and introduce some other notations.
	
	The stabilizer $S$ is a subgroup of all $n$-fold tensor products of Pauli matrices with multiplicative factors $\pm1,\pm i$ denoted by $G_n$. 
	Each Pauli matrix corresponds to a particular qubit. For instance, if $n = 5$ and the element of $G_5$ is $ZIIIX$, then operators $Z$ and $X$ correspond to the first and the last qubit, respectively. 
	More generaly, the stabilizer $S$ is some Abelian subgroup of $G_n$ {that does not include the element $-I$}. The stabilizer itself can be defined as a set of all possible compositions of operators from a finite subset of $G_n$ called a {generating set. Elements from the generating set are called generators.}
	All elements of the {generating set} should be independent (i.e., there are no generators which could be represented as a composition of some other generators) and commuting.
	We note that another useful notation is a state stabilized by a certain operator. 
	A state $|\psi\rangle$ from a Hilbert space is said to be stabilized by some operator $g\in G_n$ if $g|\psi\rangle = |\psi\rangle$.
	
	We then can define a code space using stabilizer. 
	The code space $C(S)$ is a subspace of the composite system, in which every state of the subspace is stabilized by any element of the stabilizer $S$.
	
	Below one can find examples of stabilizer codes alongside corresponding {generating sets} for which we apply the method of the following section:
	\begin{itemize}
		\item the 3-qubit repetition code with stabilizer's {generating set}
		\begin{equation}
			\langle ZZI,ZIZ\rangle.
		\end{equation}
		\item the Laflamme's 5-qubit code with stabilizer's {generating set}
		\begin{equation}
			\langle ZXXZI,XXZIZ,XZIZX,ZIZXX\rangle.\label{eq:5q_stabilizer}
		\end{equation}
		\item the Shor's 9-qubit code with stabilizer's {generating set}
		\begin{equation}
			\begin{split}
				& \langle ZZIIIIIII,IZZIIIIII,IIIZZIIII, \\
				& IIIIZZIII,IIIIIIZZI,IIIIIIIZZ, \\
				& XXXXXXIII,IIIXXXXXX\rangle.
			\end{split}
		\end{equation}
	\end{itemize}
	
	As it was mentioned above, 
	generators define a stabilizer $S$, and the stabilizer $S$ defines a code space $C(S)$. 
	We then assume that we have some state inside this subspace $C(S)$ and it is subject to some noise process. 
	For simplicity, let us assume that there is some known discrete set of error operators $\{\mathcal{E}_i\}$, which represent all possible errors that could affect a state $|\psi\rangle\in C(S)$. 
	This means that the state $|\psi\rangle$ could transform to $|\psi'\rangle = \mathcal{E}_k|\psi\rangle$, where it is not known in advance what exact error $\mathcal{E}_k$ from the set $\{\mathcal{E}_i\}$ has occurred. 
	There is a well-known result (see Ref.~\cite{NielsenChuang2000}) stating that if $\mathcal{E}_j^\dagger \mathcal{E}_k\notin N(S)-S$ for all $j$ and $k$, then $\{\mathcal{E}_i\}$ is a correctable set of errors. 
	Here $N(S)$ denotes a normalizer of the subgroup $S$ and can be defined as $N(S) = \{g\in G_n: gS = Sg\} = \{g\in G_n: gSg^{-1} = S\}$. 
	The fact that the set $\{\mathcal{E}_i\}$ is correctable essentially means that there exists a procedure that takes the noisy state $|\psi'\rangle$ and transforms it to the initial state $|\psi\rangle$ mitigating the noise.
	
	Additionally, assume that this error process can be defined as a linear combination of discrete set of operators $\mathcal{E}_i$ representing some particular errors. 
	It means that a state $|\psi\rangle$ will be transformed to a state $|\psi'\rangle = (\sum_i \alpha_i \mathcal{E}_i)|\psi\rangle$, where coefficients $\{\alpha_i\}$ are not known in advance. 
	There is another well-known result~\cite{NielsenChuang2000}, which shows that if an error is a linear combination of operators from the set $\{\mathcal{E}_i\}$ of correctable errors, then this error is correctable as well. %see Theorem 10.2 on p.438 in 
	These results can be used to prove that the mentioned Laflamme's 5-qubit and Shor's 9-qubit codes correct arbitrary single-qubit error, whereas the 3-qubit repetition code corrects bit-flip errors.
	
	For example, take the Laflamme's 5-qubit code and discrete set of errors:
	\begin{equation}
		\begin{split} 
			\{\mathcal{E}_i\} = & \{IIIII,XIIII,YIIII,ZIIII, \\
			& IXIII,IYIII,IZIII, \\
			& IIXII,IIYII,IIZII, \\
			& IIIXI,IIIYI,IIIZI, \\
			& IIIIX,IIIIY,IIIIZ\}, 
		\end{split}
	\end{equation}
	
	One can check that the condition $\mathcal{E}_j^\dagger \mathcal{E}_k\notin N(S)-S$ for all $j$ and $k$ is satisfied, were $S$ is a stabilizer defined by the {generating set} of the form~\ref{eq:5q_stabilizer}. Thus, $\{\mathcal{E}_i\}$ is a correctable set of errors.
	
	Any single-qubit error can be represented as linear combination of errors from the set $\{\mathcal{E}_i\}$. 
	For instance, arbitrary error in the second qubit $I\mathcal{E}III$ can be decomposed into the linear combination of operators from the set of correctable errors:
	
	\begin{center}
		\begin{figure}[]
			\includegraphics[scale=0.4]{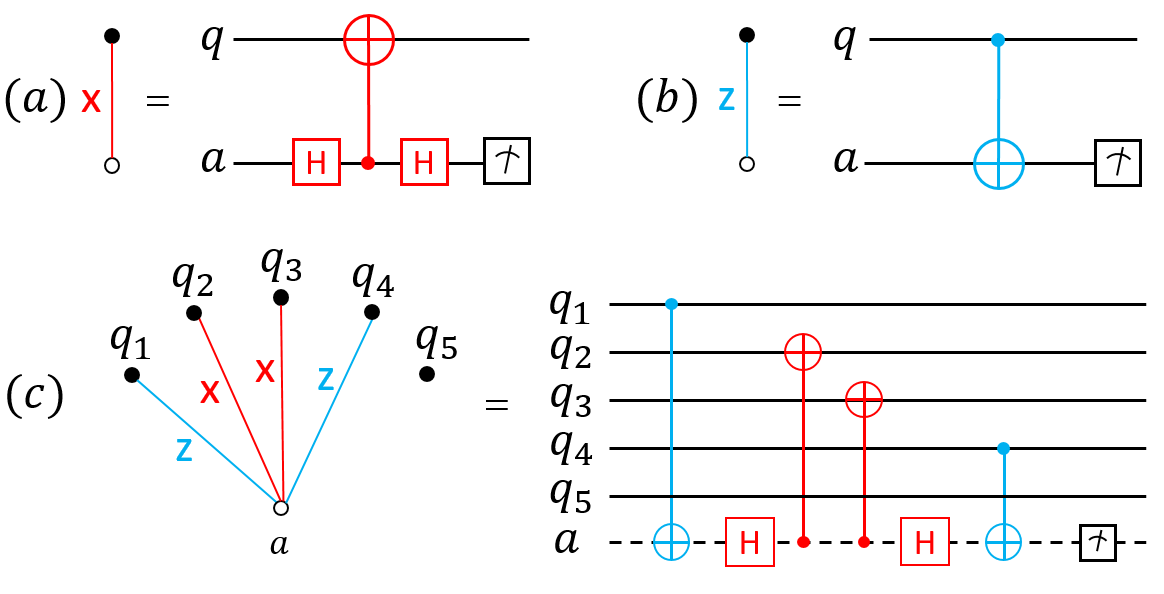}
			\vskip -3mm
			\caption{Circuits for executing measurements defined by $X$ operator in (a), $Z$ operator in (b), and $ZXXZI$ operator in (c).}
			\label{Meaurements}
		\end{figure}
	\end{center}
	
	\begin{equation}
		\begin{split}
			I\mathcal{E}III = &\alpha_I IIIII + \alpha_X IXIII + \\
			& \alpha_Y IYIII + \alpha_Z IZIII.
		\end{split}
	\end{equation} 
	It essentially means that any single-qubit error is correctable. 
	The same argumentation can be applied to prove that any single-qubit error in the nine-qubit code is correctable. 
	In the case of the 3-qubit code, the discrete set of correctable errors is as follows:
	\begin{equation}
		\{\mathcal{E}_i\} = \{III,XII,IXI,IIX\}.
	\end{equation} 
	
	The procedure that realizes transformation of the noisy state $|\psi'\rangle$ into the initial state $|\psi\rangle$ has two steps: 
	(i) at the first step syndrome measurements are performed and (ii) at the second step unitary transformation for a correction conditional on syndrome measurement results is found. 
	The detailed description of the process of finding a correction for stabilizer codes of our interest can be found in Sec.~\ref{sec:decoding}. 
	The rest of this section is devoted to the description of performing syndrome measurements.
	
	It can be shown that syndrome measurements should be defined by operators from the {generating set} of the stabilizer $S$. 
	To make measurements that are defined by generators, circuits with ancilla qubits can be used. 
	{Ancilla, which is taken in a known computational basis state, after applying the designed circuit, is measured in the standard computational basis.It implicitly leads to a required measurement defined by a generator.}
	For example, if we are interested in the measurement for the data qubit $q$ defined by the operator $X$ using ancilla qubit $a$, then the circuit Fig.~\ref{Meaurements}(a) is used for such a measurement. 
	The circuit illustrated in Fig.~\ref{Meaurements}(b), in turn, is used for the measurement defined by $Z$. 
	We place the detailed description of the workflow of circuits (a) and (b) in the Appendix~\ref{sec:App1}.
	
	For multiqubit circuits, the measurement defined by a tensor product of matrices $I$, $X$ and $Z$ can be realized in a similar fashion: 
	one just needs to combine corresponding gates using one common ancilla and make a measurement of the ancilla in the end of the circuit. 
	For instance, the circuit for the measurement defined by the operator $ZXXZI$ in the five-qubit system is presented in Fig.~\ref{Meaurements}(c).
	
	If stabilizer code is defined by a stabilizer $S$ with the {generating set} $\{g_1,g_2,...,g_k\}$, then to obtain syndrome measurements one needs to repetitively make measurements defined by every element from the {generating set}. 
	With the use of syndrome measurements, it is possible to come up with a unitary transformation that corrects for a noise process that occurred with logical qubits during measurements.
	
	\subsection{CNS gate}
	
	Superconducting qubits have native two-qubit operations which are easier to realize than standard two-qubit operations~\cite{Siewert2003}. 
	One such native operation is iSWAP gate:
	
	\clearpage
	
	\begin{center}
		\begin{figure}[]
			\includegraphics[scale=0.45]{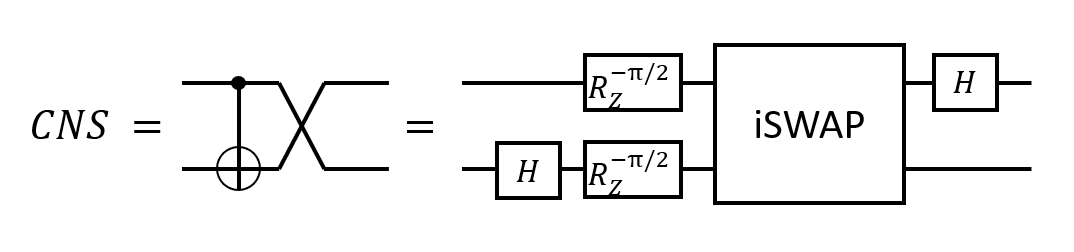}
			\vskip -4mm
			\caption{Correspondence between two-qubit gates iSWAP and CNS. Standard notations for Pauli rotations and traditional gates are used.}
			\label{CNS}
		\end{figure}
	\end{center}
	
	\begin{equation}
		\text{iSWAP} = \begin{pmatrix}
			1 & 0 & 0 & 0\\
			0 & 0 & i & 0\\
			0 & i & 0 & 0\\
			0 & 0 & 0 & 1\\
		\end{pmatrix}.
	\end{equation}
	This native operation can be supplemented by simple one-qubit operations to get CNS transformation illustrated in Fig.~\ref{CNS}. 
	CNS is equivalent to the composition of CNOT and SWAP gates, and this allows us to replace CNOT with CNS in circuits realizing syndrome measurements.
	
	\section{Single-ancilla realization}\label{sec:single-ancilla}
	
	Our method is inspired by two important peculiarities in experimental realizations of superconducting qubits.
	The first is that some patterns of connectivity between qubits are easier to implement than the others. 
	Specifically, connectivity of existing experimental platforms is mostly pairwise, i.e. all two-qubit interactions are allowed between neighboring qubits~\cite{Oliver2020}. 
	Therefore, in this work we focus on the circular pattern of connectivity between qubits.
	The second aspect is that superconducting qubits have native two-qubit operations, which are easier to realize than standard two-qubit operations~\cite{Siewert2003}. 
	Using native operation iSWAP and single-qubit operations, it is possible to realize CNS gate that is equivalent to CNOT and SWAP gates. 
	Thus, the CNS gate could be used to replace CNOT gates in the circuit for syndrome measurements.
	
	Below we provide circuits for realizations of the 3-qubit repetition code and the Laflamme's 5-qubit code. 
	A circuit for the seminal Shor's nine-qubit code is provided in Appendix~\ref{sec:nine}. 
	We note that in our realizations, first, all the circuits use only interactions of circular connectivity pattern and a single ancilla. 
	Second, all the operations in 3- and 5-qubit codes are either standard one-qubit or CNS two-qubit gates. 
	The vast majority of operations in the nine-qubit code are either standard one-qubit or CNS two-qubit gates, and there are few CNOT-gates.
	
	\begin{center}
		\begin{figure}[]
			\includegraphics[scale=0.375]{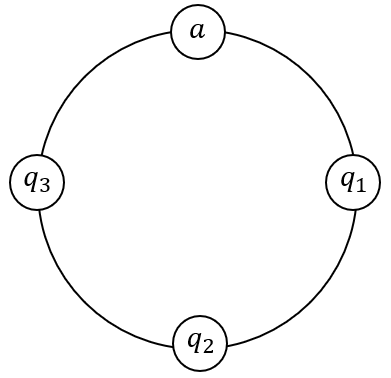}
			\vskip-2mm
			\caption{Circular connectivity pattern for 4 qubits.}
			\label{Chain3}
		\end{figure}
	\end{center}
	
	\begin{center}
		\begin{figure*}[t]
			\includegraphics[scale=0.66]{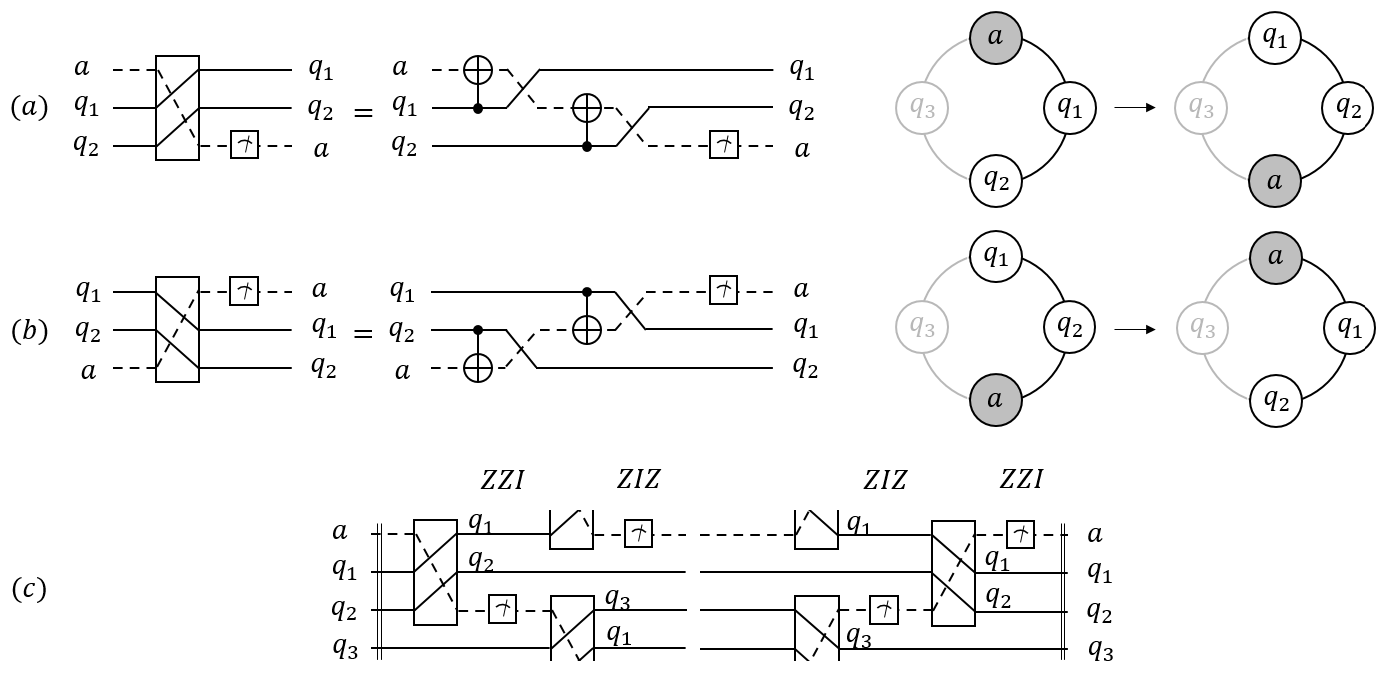}
			\centering\rule{.7\textwidth}{0\textwidth}
			\caption{Realization of the 3-qubit code: (a) ``Riffle'' block for a measurement defined by the operator $ZZI$. 
				(b) ``Reverse riffle'' block for a measurement defined by the operator $ZZI$. 
				(c) The circuit realizing 1 cycle of consecutive syndrome measurements for the 3-qubit repetition code.}
			\label{Schemes_3q}
		\end{figure*}
	\end{center}
	
	\begin{center}
		\begin{figure}[!ht]
			\includegraphics[scale=0.33]{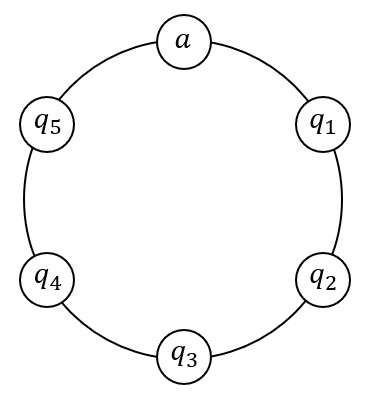}
			\caption{Circular connectivity pattern for 6 qubits.}
			\label{Chain5}
		\end{figure}
	\end{center}
	
	\vspace{-0.7in}
	
	\subsection{3-qubit code}
	
	Let us assume that we want to encode a logical {qubit} into three physical qubits, and we have 1 ancilla qubit. 
	Suppose additionally, that all 4 qubits are connected in the circular pattern (Fig.~\ref{Chain3}). 
	This means that two-qubit interactions are allowed between nearest-neighbor qubits only. 
	We note that it is not important in which place exactly the ancilla qubit $a$ is located. 
	Also we assume that data qubits $\{q_k\}_{k=1}^3$ are ordered by their index number.
	
	We can use two types of gate blocks, which we refer to as ``riffles''. 
	These blocks essentially realize a measurement defined by some generator, depending on what qubits they are applied to.
	Riffles are represented in Fig.~\ref{Schemes_3q}(a) and Fig.~\ref{Schemes_3q}(b). 
	When applied to the qubits $a, q_1$ and $q_2$, riffle and reverse riffle are two different realizations of the same measurement defined by the element $ZZI$.
	
	We note that these riffle gates exploit only interactions between neighboring qubits and only CNS gates for two-qubit interactions. 
	Additionally, after we apply the riffle, it moves the information from one physical qubit to another. 
	This takes place since CNS gate contains SWAP operation as its part. 
	Note also that the pattern of such a movement is schematically captured by the riffle's notation. 
	One can see that the scheme from the bottom looks like an inverse version of the scheme from the top, 
	but in fact, due to measurements at the end of circuits and ordering of operations these schemes are not inverse versions of each other.
	
	The circuit realizing measurements by elements $ZZI$ and $ZIZ$ is presented in Fig.~\ref{Schemes_3q}(c).
	It consists of two parts: in the first part two riffle gates are applied and syndromes for $ZZI$ and $ZIZ$ elements are measured, 
	and in the second part two reverse riffle gates are applied and syndromes for $ZIZ$ and $ZZI$ elements are measured. 
	We note that the scheme can be repetitively applied to get the syndrome for the 3-qubit repetition code.
	There are also two modifications of the given scheme (see Appendix~\ref{sec:Modifications}).

	\begin{center}
		\begin{figure*}[!ht]
			\includegraphics[scale=0.68]{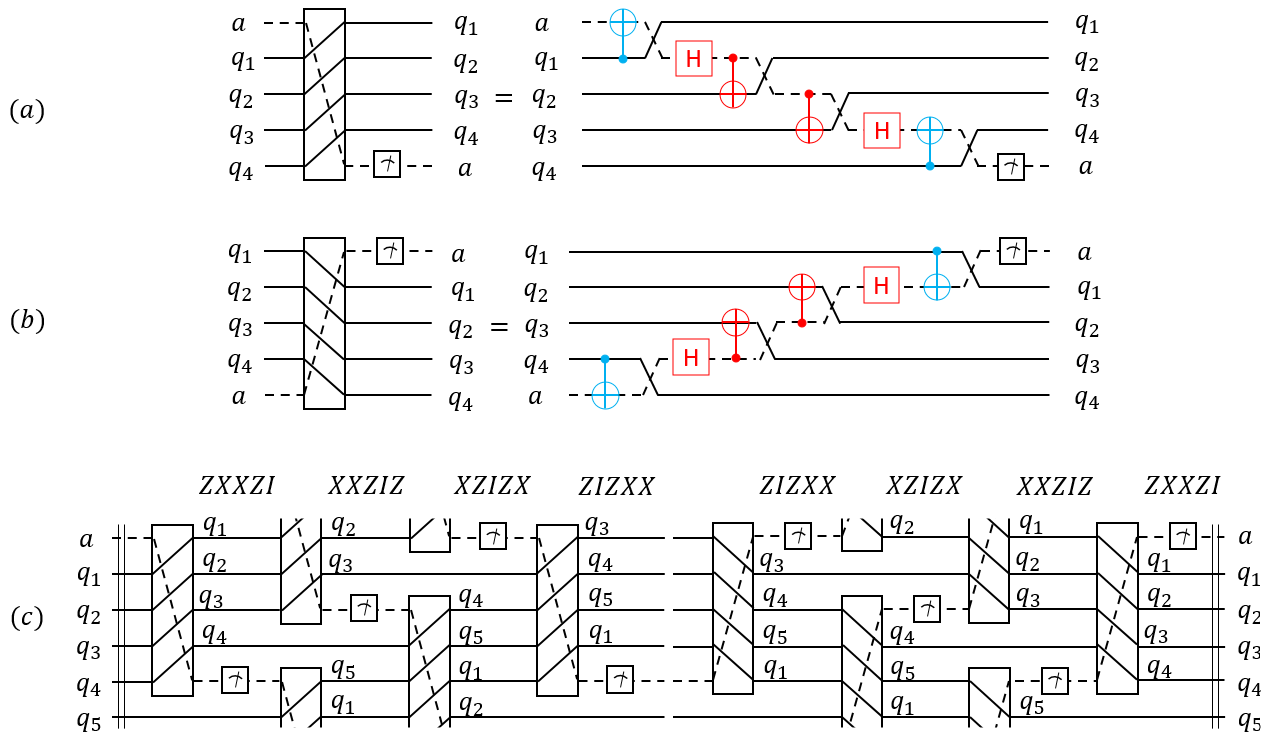}
			\caption{Realization of the Laflamme's 5-qubit code: 
				(a) ``Riffle'' block for a measurement defined by the operator $ZXXZI$. (b) ``Reverse riffle'' block for a measurement defined by the operator $ZXXZI$. (c) The circuit realizing consecutive syndrome measurements for the Laflamme's 5-qubit code.}
			\label{Schemes_5q}
		\end{figure*}
	\end{center}
	
	\subsection{Laflamme's 5-qubit code}
	
	The same ideas can be applied to get the realization of the Laflamme's 5-qubit code. 
	Assume we have 6 qubits connected in a circular chain as illustrated in Fig.~\ref{Chain5}: 
	There are 5 data qubits $\{q_k\}_{k=1}^5$ ordered by their index number and 1 ancilla qubit.
	
	The two types of riffle gates that are used in the construction of a code exploit only interactions between neighboring qubits and only CNS gates for two-qubit interactions, as illustrated in Fig.~\ref{Schemes_5q}(a) and Fig.~\ref{Schemes_5q}(b). 
	If entries of these gates are as denoted in the picture, then both gates realize a measurement defined by the operator $ZXXZI$. 
	Any other operator from the generating set of the 5-qubit code's stabilizer can be measured with these gates by proper choice of entry-qubits. 
	Gates denoted by blue (light gray) color correspond to $Z$-entries in the generator, 
	and gates denoted by red (dark gray) color correspond to $X$-entries.
	
	The scheme in Fig.~\ref{Schemes_5q}(c) is the realization of measurements of generators of the Laflamme's 5-qubit code. 
	In the first part  $ZXXZI,XXZIZ,XZIZX,ZIZXX$ are measured, and in the second part the same set of generators is measured in the reverse order: $ZIZXX,XZIZX,XXZIZ,ZXXZI$.
	
	We note that the idea of using {\sf CNS} gates and linear connectivity for the five-qubit code is also presented in Ref.~\cite{Siewert2003},
	where authors provided a circuit consisting of 5 data- and 5 ancilla-qubits. 
	A notable realization of the Laflamme's 5-qubit code structure that makes use of 5 data qubits, 5 ancilla qubits, and iSWAP gates can be found in Ref.~\cite{Ustinov2022}.
	In the present work, we provide the circuit using only 1 ancilla qubit and generalize this approach to other stabilizer codes.
	{Additionally, it is worth mentioning that a somewhat similar single-ancilla measurement structure for the four-qubit code exploiting chain connectivity pattern was suggested in~\cite{De_2013}.}
	
	\section{Generalization of the method}\label{sec:generalization}
	
	\subsection{Circular connectivity}
	
\begin{center}
		\begin{figure}[!ht]
			\includegraphics[scale=0.375]{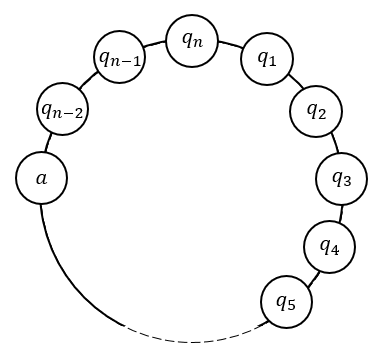}
			\vskip-4mm
			\caption{General circular connectivity pattern for $n+1$ qubits.}
			\label{GeneralChain}
		\end{figure}
	\end{center}

	Suppose we want to encode a logical {qubit} into $n$ physical qubits, and we have 1 ancilla qubit. 
	We additionally assume that all $n+1$ qubits are organized in the circular pattern, as it is shown in Fig.~\ref{GeneralChain}. 
	This means that two-qubit interactions are allowed between neighboring qubits only. 
	As for now, it is not important where exactly ancilla qubit $a$ is placed, but we assume that data qubits $\{q_k\}_{k=1}^n$ are ordered by their index number.
	
	Suppose that $g_i$ denotes some element of a {generating set} and all non-trivial entries (i.e. matrices $X$ and $Z$) in $g_i$ are organized in one block. 
	This means that non-trivial entries correspond to neighboring qubits (not counting ancilla qubit). 
	Let us denote by $L_{i}$ an index of the place corresponding to the first non-trivial entry in the block, and by the $R_{i}$ an index of the place corresponding to the last non-trivial entry in the block.
	
	For instance, the element $g_1 = IZXXZIII...III$ has non-trivial entries corresponding to neighboring qubits $\{q_2,q_3,q_4,q_5\}$. 
	We see that the first non-trivial entry is at the place $2$ and the last non-trivial entry at the place $5$. Thus, $L_1 = 2$ and $R_1 = 5$. 
	The element $g_2 = XZIIIIII \dots IZX$ has non-trivial entries corresponding to neighboring qubits $\{q_{n-1},q_n,q_1,q_2\}$. 
	The first non-trivial entry is at the place $n-1$, where $n$ is the amount of data qubits, and the last non-trivial entry at the place $2$. Thus, $L_2 = n-1$ and $R_2 = 2$.
	
	Consider stabilizer codes with two restrictions on the {generating set} of a stabilizer:
	\begin{itemize}
		\item all non-trivial entries (i.e. matrices $X$ and $Z$) in every generator are organized in one block;
		\item generators can be ordered in such a sequence $\{g_i\}$ that for any $i\in[1,n-1]$ one of the following requirements is satisfied:
		\begin{enumerate}
			\item $R_i = \begin{cases}
				L_{i+1} - 1 &\text{if $L_{i+1} > 1$}\\
				n &\text{if $L_{i+1} = 1$}
			\end{cases}.$
			
			$\text{Example:}\quad\begin{matrix} \cline{4-6}
				g_i = & I & I & \multicolumn{1}{|c}{Z} & Z & \multicolumn{1}{c|}{Z} & I & I & I & I\\ \cline{4-9}
				g_{i+1} = & I & I & I & I & I & \multicolumn{1}{|c}{Z} & Z & \multicolumn{1}{c|}{Z} & I\\ \cline{7-9}
			\end{matrix}$
			\item $R_i = L_{i+1}.$
			
			$\text{Example:}\quad\begin{matrix} \cline{4-6}
				g_i = & I & I & \multicolumn{1}{|c}{Z} & Z & \multicolumn{1}{c|}{Z} & I & I & I & I\\ \cline{4-8}
				g_{i+1} = & I & I & I & I & \multicolumn{1}{|c}{Z} & Z & \multicolumn{1}{c|}{Z} & I & I\\ \cline{6-8}
			\end{matrix}$
			\item $R_i = \begin{cases}
				L_{i+1} + 1 &\text{if $L_{i+1} < n$}\\
				1 &\text{if $L_{i+1} = n$.}
			\end{cases}.$
			
			$\text{Example:}\quad\begin{matrix} \cline{4-6}
				g_i = & I & I & \multicolumn{1}{|c}{Z} & Z & \multicolumn{1}{c|}{Z} & I & I & I & I\\ \cline{4-7}
				g_{i+1} = & I & I & I & \multicolumn{1}{|c}{Z} & Z & \multicolumn{1}{c|}{Z} & I & I & I\\ \cline{5-7}
			\end{matrix}$
		\end{enumerate}
	\end{itemize}
	
	\begin{center}
		\begin{figure*}[!ht]
			\includegraphics[scale=0.66]{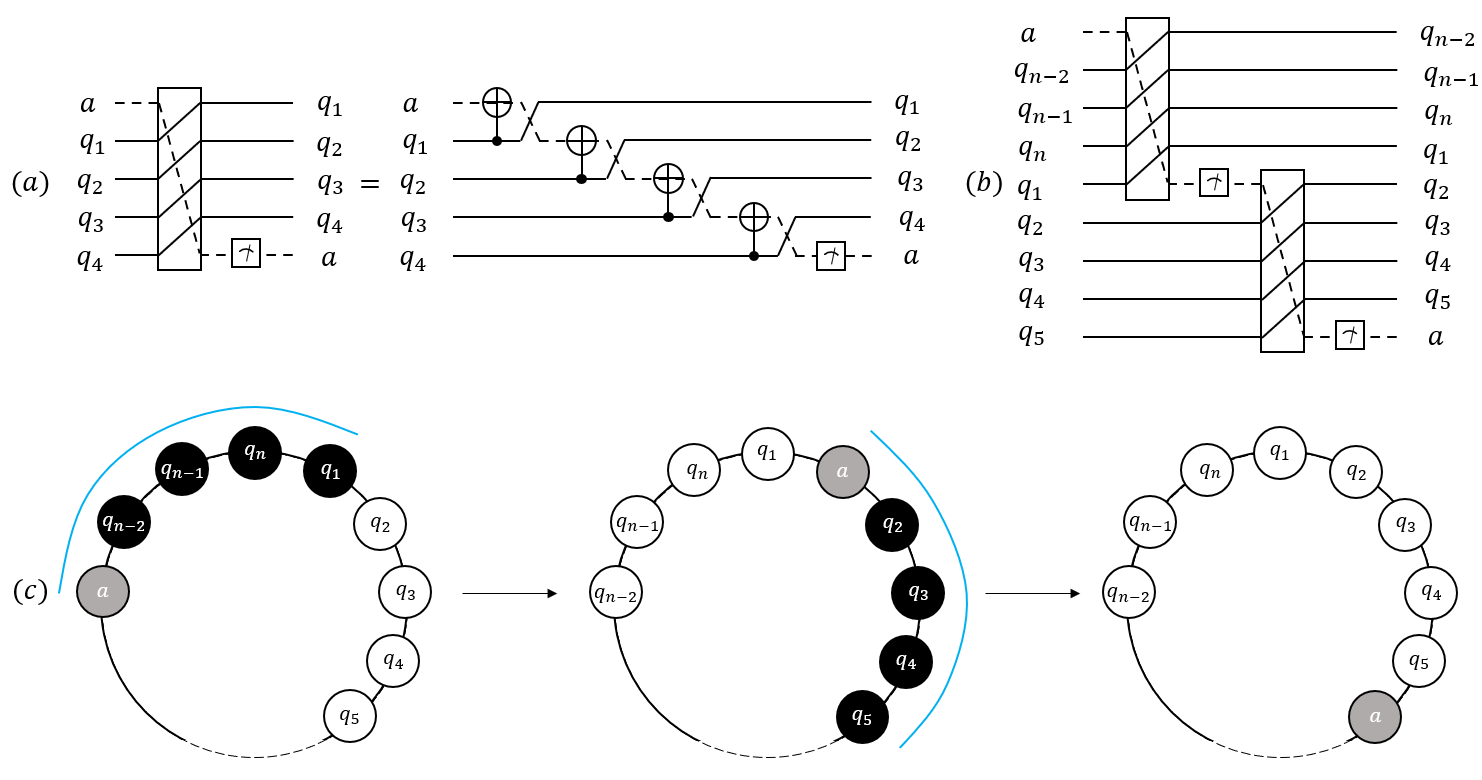}
			\caption{(a) ``Riffle'' block realizing a measurement. 
				(b) Two ``riffle'' blocks are combined to make use of a single ancilla. 
				(c) ``Movement'' of ancilla as the direct effect of applying the two ``riffle'' blocks.}
			\label{GeneralExample}
		\end{figure*}
	\end{center}
	
	Let us refer the class of such stabilizer codes ``neighboring blocks'' stabilizer codes. 
	As can be seen from examples, Condition 1 corresponds to the case of ``adjacent'' non-trivial blocks, Condition 2 corresponds to the case of non-trivial blocks ``overlapping'' by one element, 
	and condition 3 corresponds to the case of non-trivial blocks ``overlapping`` by two elements. 
	We note that ``overlapping'' or ``adjacency'' is considered only for the last element of the non-trivial block in $g_i$ and the first element of the non-trivial block in $g_{i+1}$, 
	and there are no restrictions on other elements of blocks for every given pair $\{g_i,g_{i+1}\}$. 
	Having said that, note that if $\{g_1,g_2\} = \{ZXXZI,XXZIZ\}$, then condition 1 is satisfied, since $R_1 = 4 = L_2-1$.
	
	3-qubit repetition code, Laflamme's 5-qubit code, and Shor's 9-qubit code are examples of ``neighboring blocks'' stabilizer codes. 
	This class is of particular interest to us, because if there is such a structure, then well-defined procedure could be applied to build a circuit realizing measurements defined by generators with circular pattern of connectivity and a single ancilla.
	
	To understand the procedure, we can consider the case when we have some ``neighboring blocks'' stabilizer code for which Condition 1,
	\begin{equation}
		R_i = \begin{cases}
			L_{i+1} - 1 &\text{if $L_{i+1} > 1$}\\
			n &\text{if $L_{i+1} = 1$},
		\end{cases}
	\end{equation}
	is satisfied for $i = 1$. 
	For instance, let the ordered set of generators be $\{g_1,g_2,...\} = \{ZII\dots IIZZZ,\;IZZZZII \dots II,\;\dots\}$. It is clear that $R_1 = 1 = L_2 - 1$.
	
	Consider a circuit with two consecutive blocks of the form given in Fig.~\ref{GeneralExample}(a) applied to a system of qubits illustrated in Fig.~\ref{GeneralChain}: 
	we apply the first block to qubits $\{a,q_{n-2},q_{n-1},q_{n},q_{1}\}$ and then the second block to qubits $\{q_1,q_2,q_3,q_4,q_5\}$. 
	This circuit corresponds to the ``movement'' of information in the chain as it is shown in Fig.~\ref{GeneralExample} (b) and (c).
	
	During this movement, two stabilizer elements are measured: $ZIIIIII\dots IZZZ$ and $IZZZZII \dots IIII$. 
	By continuing a sequence of blocks, measurements for other elements of the stabilizer will be realized, until all the generators are measured.
	
	\begin{center}
		\begin{figure*}[!ht]
			\includegraphics[scale=0.4]{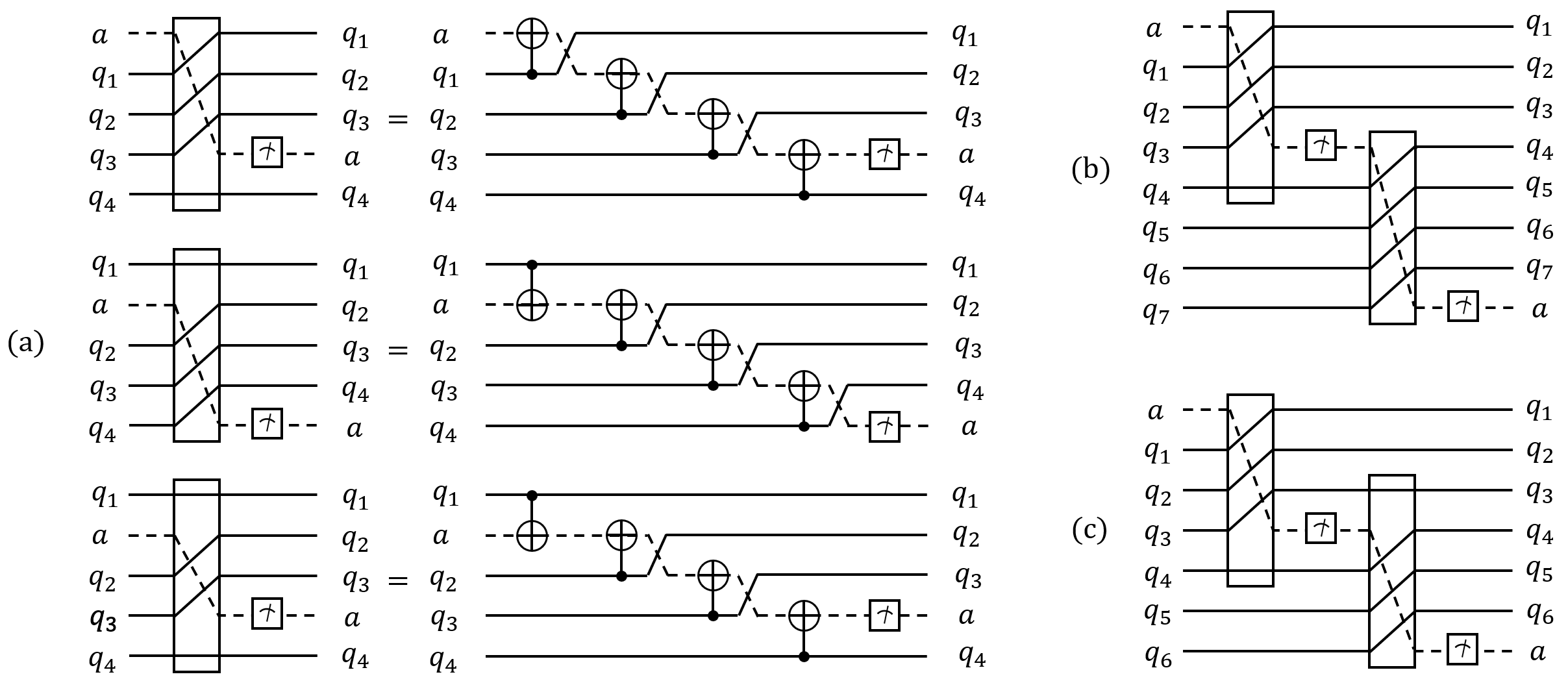}
			\caption{(a) ``Riffle'' blocks with CNOT gates either at the end or at the beginning of the block. 
				(b) Circuit realizing measurements defined by stabilizers with 1 ``overlapping`` element: $ZZZZIII$ and $IIIZZZZ$. 
				(c) Circuit realizing measurements defined by stabilizers with 2 ``overlapping`` element: $ZZZZII$ and $IIZZZZ$.}
			\label{OverlappingSchemes}
		\end{figure*}
	\end{center}
	
	This procedure can be generalized straight-forwardly for any pair of generators $\{g_i,g_{i+1}\}$ with ``adjacent`` blocks, since one can just construct riffle block which will perform measurement defined by corresponding stabilizer. 
	To account for $Z$-entry, CNS gate should be applied. 
	CNS is equivalent to CNOT and SWAP gates, thus, if control-qubit for CNOT corresponds to the place of $Z$-entry, 
	then resulting measurement operator will include this $Z$-entry. 
	As can be seen from Fig.~\ref{Meaurements}(b), SWAP-gate does not change how circuit elements work if proper qubit is measured.
	The same can be said about $X$-entry: one can just add Hadamard gates to reproduce element as in Fig.~\ref{Meaurements}(a), but with CNS instead of CNOT gate. 
	As an illustration of how this riffle gate performing proper measurement should be organized, we consider Fig.~\ref{Meaurements}(c) and Fig.~\ref{Schemes_5q}(a) 
	and note that both these circuits are realizing the same measurement defined by the operator $ZXXZI$.
	
	To cope with ``overlapping'' pairs of generators, we introduce additional ``riffle with CNOTs'' types of measurement blocks. 
	They are provided in Fig.~\ref{OverlappingSchemes}(a) for the case of 4 non-trivial $Z$-entries, 
	but it is easy to construct these types of riffle measurement blocks for any number of non-trivial entries. 
	The only difference between ``riffle'' and ``riffle with CNOT'' types is that either the first or the last (or both) CNS gate in the riffle is replaced with CNOT gate. 
	Examples of realizing measurements defined by pair of operators with 1 and 2 overlapping elements are provided in Fig.~\ref{OverlappingSchemes} (b) and (c).
	
	After we organized generators in a sequence of ``neighboring'' elements, two strategies are possible for repetitive measurements of the stabilizer's elements.
	The first strategy is that we can make measurements of the same generators of the stabilizer by applying ``reverse'' versions of the blocks from the first full cycle.
	Alternatively, we can continue applying the same set of blocks to the same physical qubits after the full cycle of the generators' measurements if this is possible. 
	It might be the case that another generator of the same stabilizer is measured, and using additional classical decoding of a syndrome we still have a realization of required code.
	
	Note that the method can produce CNOT gates depending on the type of riffle gates used, as can be seen, for instance, in Fig.~\ref{OverlappingSchemes}. 
	Thus, for more efficient realizations it is better to use only riffle gates without CNOT elements if possible.
	
	{We note that there is a large class of quantum cyclic codes (for a review, see Refs.~\cite{https://doi.org/10.48550/arxiv.1007.1697,https://doi.org/10.48550/arxiv.quant-ph/9608006}), including a [[13,1,5]] cyclic code in~\cite{Kovalev_2011} similar to the 5-qubit code. These codes might have some similarity with "neighboring blocks" stabilizer codes, but we leave this for further research.}

	\begin{center}
		\begin{figure}[!ht]
			\includegraphics[scale=0.7]{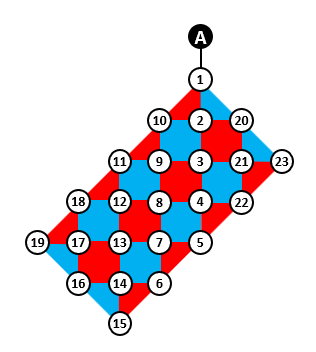}
			\vskip-4mm
			\caption{{Example of surface code layout. Blue (light gray) faces are associated with X-elements of the stabilizer and red (dark gray) faces are associated with Z-elements of the stabilizer. A black qubit is an ancilla.}}
			\label{SurfaceCode}
		\end{figure}
	\end{center}
	
	\subsection{Near-neighbour connectivity}\label{sec:SurfaceCode}
	
	{If we relax the circular connectivity condition and allow only near-neighbour interactions, there is a way to realize all the required measurements for surface code in sequential manner using elements that are similar to riffle gates in Fig.~\ref{OverlappingSchemes}. Let us consider the particular example of surface code that is schematically given in Fig.~\ref{SurfaceCode}. This scheme contains 23 data qubits and 1 ancilla-qubit. Each blue (light gray) face corresponds to X-element of stabilizer and red (dark gray) face corresponds to Z-element of stabilizer. For example, blue (light gray) face between qubits 10, 2, 3 and 9 corresponds to the element $X_{10}X_2X_3X_9$. Stabilizer's {generating set} contains of 22 elements, thus, this layout encodes 1 logical qubit.}
	
	\begin{center}
		\begin{figure*}[!ht]
			\includegraphics[scale=0.8]{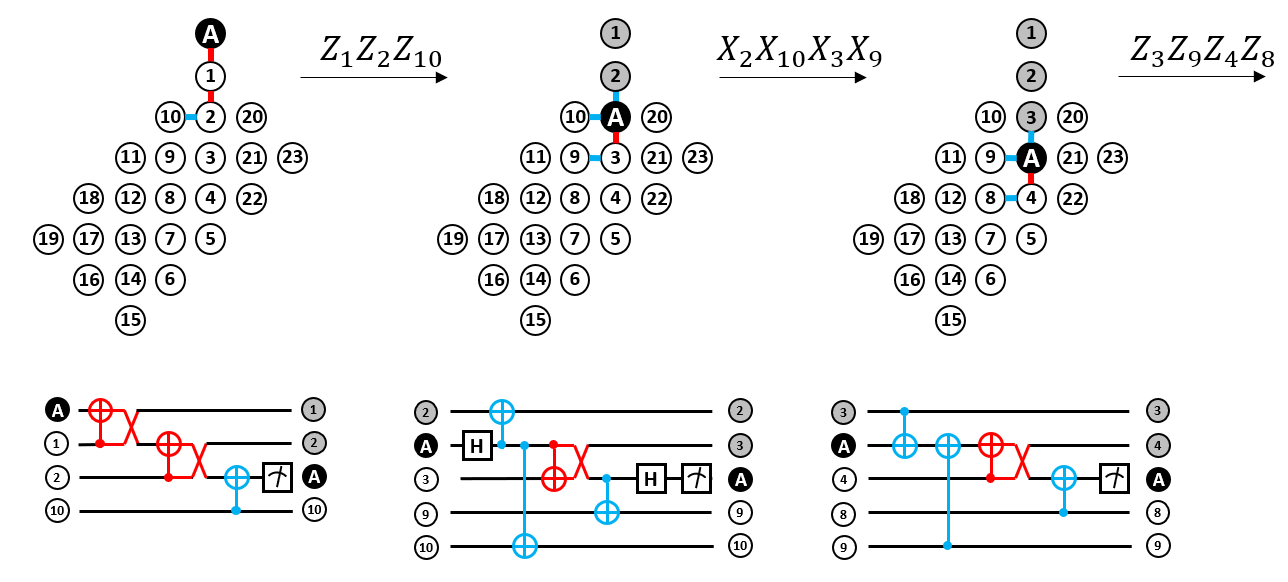}
			\caption{{The first three measurements of Surface code's stabilizer and corresponding circuit elements. Red (dark gray) edge corresponds to CNS gate, and blue (light gray) edge corresponds to CNOT gate. The numeration of qubits has resemblance with the order of applying gates.}}
			\label{SurfaceCode_1_3}
		\end{figure*}
	\end{center}
	
	{Numerical order of data qubits reflects the order of applying gates in a sequential scheme. The first 3 steps of the scheme are given in Fig.~\ref{SurfaceCode_1_3}, and all the remaining steps are provided in Appendix~\ref{sec:surfacecode_1_22_steps}.}
	
	{This example suggests that it is possible to generalize this result not only to surface codes, but to other codes with near-neighbour structure of stabilizers, such as considered in Ref.~\cite{https://doi.org/10.48550/arxiv.2203.16486} or in Ref.~\cite{Fowler_2012}. This generalization is a subject of further research.}
	
	\section{Decoding}\label{sec:decoding}
	
	When all syndrome measurements are performed, classical post-processing should be applied to the syndrome to obtain a set of transformations for proper correction of the stored state. 
	There is a well-known approach for efficient decoding in the 3-qubit code called minimum weight perfect matching. 
	However, this approach relies on simultaneous measurements defined by the generators, whereas our scheme exploits consecutive measurements. 
	We first briefly describe minimum weight perfect matching for simultaneous measurements in repetition codes, then provide adaptation of this algorithm for consecutive measurements in our scheme for repetition code, 
	and, finally, we describe the adaptation of the algorithm for the case of the Laflamme's 5-qubit code.
	Results of simulations using the provided adaptation technique for decoding are provided in the next section.
	
	\subsection{Minimum weight perfect matching for simultaneous measurements in the repetition code}
	
	Here we describe and illustrate the well-known algorithm for decoding bit-flip errors from the syndrome consisting of measurements defined by stabilizer set $\langle ZZI,ZIZ\rangle$. 
	The description of the same algorithm can be found in Ref.~\cite{Kelly2015}, but in order to further introduce a modification of the algorithm for our schemes, 
	we would need proper notations.
	For this purpose, we repeat the description here, but using our own notations.
	
	\begin{center}
		\begin{figure*}[!ht]
			\includegraphics[scale=0.7]{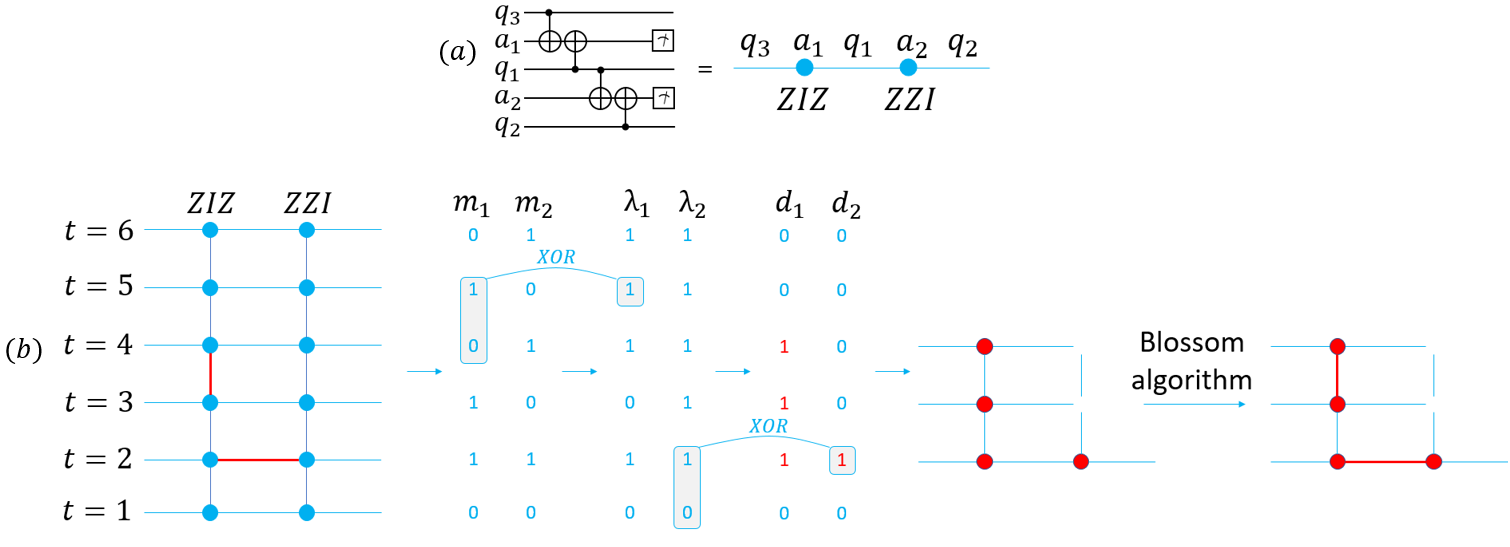}
			\caption{Decoding procedure. In (a) we introduce notation for the decoding scheme. In (b) we illustrate steps of the decoding algorithm.}
			\label{RepetitonGraph_DecodingAlgorithm}
		\end{figure*}
	\end{center}
	
	Suppose we have 5 qubits, as it is shown in Fig.~\ref{RepetitonGraph_DecodingAlgorithm}(a): 3 of them ($q_1,q_2$ and $q_3$) store data, and 2 of them ($a_1$ and $a_2$) are ancilla qubits used for measurements defined by the generators. 
	Let us then denote data qubits by edges and ancilla qubits by nodes. 
	For the process of repetitive simultaneous measurements defined by stabilizers $ZIZ$ with ancilla qubit $a_1$ and $ZZI$ with ancilla qubit $a_2$ we construct a graph. 
	Time goes from the bottom to the top, and every horizontal element corresponds to the stance of the five qubits at the moment $t$. 
	Vertical edges are used for detecting errors in ancilla qubits, as will be shown later.
	
	{We should mention that ancilla-measurements presented here are assumed to be quantum non-destructive (QND). This means that after a measurement, ancilla-qubit preserves the measured state and it is not subject to reinitialization. We describe the decoding procedure for non-destructive measurements. Explanation of differences between non-destructive ancilla-measurements and measurements with reinitializations can be found in the end of the paragraph.}
	
	Ancilla qubits are initialized in the state $|0\rangle$. 
	There are only two possible states of ancilla qubits after measurement: $|0\rangle$ and $|1\rangle$. 
	If data qubits are initialized as $|1\rangle$ and ancilla qubit is flipping during measurements, i.e. constantly changes state from $|0\rangle$ to $|1\rangle$ and vice versa, then it indicates that one of the corresponding data qubits have had a bit-flip error. 
	If ancilla qubit is stable, i.e. always $|0\rangle$ or always $|1\rangle$, then it indicates that corresponding data qubits have similar states, i.e. either no error has occurred or a bit-flip error occurred on both data qubits. 
	Results of non-destructive measurements of ancilla qubits are used to reconstruct error processes during the measurements. 
	We assume that only two types of errors are possible: a bit-flip error ($X$-error) in one of the data qubits or a bit-flip error ($X$-error) during the measurement of an ancilla qubit which leaves ancilla in the state that is opposite to the correct one.
	
	The algorithm relies on finding ``detection events'' and then building subgraph which tells us the most probable candidates for errors. 
	To illustrate the decoding algorithm, suppose data qubit $q_2$ was subject to bit-flip error on the step $t = 2$ and ancilla qubit was subject to bit-flip error on the step {$t = 3$}. 
	Let us denote by $m_j^t$ result of measurement of ancilla qubit $a_j$ on the step $t$: $m^t_j = 0$ if $a_j$ is in the state $|0\rangle$ and $m^t_j = 1$ if $a_j$ is in the state $|1\rangle$. 
	Then we can calculate values $\lambda^t_j = m^t_j\oplus m^{t-1}_j$. If $\lambda^t_j = 0$, then it indicates that ancilla has not flipped from step $t-1$ to step $t$, and if $\lambda^t_j = 1$, then it indicates that ancilla has flipped from step $t-1$ to step $t$. 
	Then we can calculate values $d^t_j = \lambda^t_j\oplus \lambda^{t-1}_j=m_j^t\oplus m_j^{t-2}$. If $d^t_j = 0$, then behavior of ancilla in terms of flipping/not flipping has not changed from step $t-1$ to step $t$ and we do not have detection event. 
	If $d^t_j = 1$, then behavior of ancilla in terms of flipping/not flipping has changed from step $t-1$ to step $t$ and we have detection event. 
	The subgraph is constructed from nodes corresponding to detection events and all possible connections of these nodes, as shown in Fig.~\ref{RepetitonGraph_DecodingAlgorithm}(b). 
	The weight of each connection is equal to the number of edges used to form this connection. 
	
	Finally, Blossom algorithm is applied to the subgraph to find minimum weight perfect matching, 
	i.e. the set of non-adjacent edges with the following properties: every node of the subgraph is matched, and the total weight of all edges in such a matching is minimal among all possible matchings that include every node of the subgraph.
	
	Horizontal edge in the resulting matching is interpreted as a bit-flip error in the corresponding data qubit. 
	Vertical edge in matching is interpreted as a bit-flip error in the corresponding ancilla qubit. 
	The given algorithm is efficient since one can efficiently calculate indicators of detection events $d^t_j$ and then apply an efficient Blossom algorithm to get the set of the most probable errors.
	
	{The decoding procedure for the case when ancilla-qubits are reinitialized to the state $|0\rangle$ after each measurement is slightly simpler. Note that flipping pattern in non-destructive measurement is equivalent to measuring $|1\rangle$ while using reinitialization. Analogously, absence of flipping is equivalent to measuring $|0\rangle$ with reinitialization. Thus, one difference in the decoding procedure is assigning to $\lambda_j^t$ corresponding measurement result instead of calculating this value with $m_j^t$. Another difference is that non-desctructive measurements require a couple of additional rounds of measurements to correctly identify all the values, as described in Ref.~\cite{Kelly2015}.}
	
	\begin{center}
		\begin{figure*}[!ht]
			\includegraphics[scale=0.69]{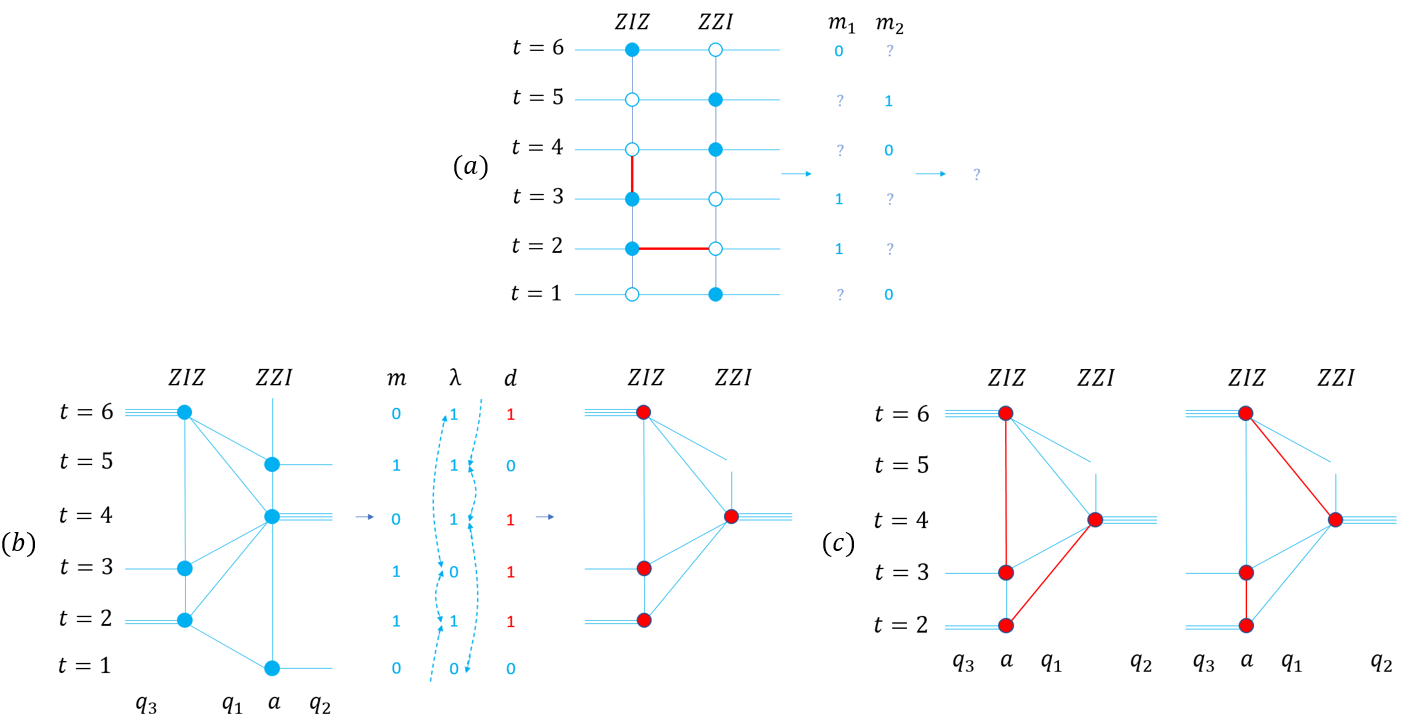}
			\caption{(a) Graph representing information from consecutive measurements. 
				Due to lacking information, usual algorithm based on minimum perfect matching for simultaneous measurements would not work.
				(b) Deformed graph representing information from consecutive measurements. 
				This graph can be used to decode errors. 
				(c) Two error patterns with the same weight after decoding.}
			\label{RepetitonGraphProblems_DecodingAlgorithm}
		\end{figure*}
	\end{center}
	
	\subsection{Minimum weight perfect matching for consecutive measurements in the repetition code}
	
	Here we provide an adaptation of the minimum-weight perfect matching algorithm for the case of consecutive measurements exploited in the realization of the 3-qubit code with linear connectivity and a single ancilla. 
	Recall that the circuit illustrated in Fig.~\ref{Schemes_3q}(c) realizes measurements in the ``forward-backward'' consecutive manner: $ZZI$, $ZIZ$, $ZIZ$, $ZZI$, etc. 
	If we use the same graph as for simultaneous measurements, we find that information about the measurements is only partial: 
	we do not know the values of $m^1_1,m^2_2,m^3_2,m^4_1$ and $m^5_1$, as depicted in Fig.~\ref{RepetitonGraphProblems_DecodingAlgorithm}(a). 
	Thus, we cannot calculate properly detection events $d^t_j$ using the known algorithm. 
	Additionally, notice that one ancilla is used for both generators $ZZI$ and $ZIZ$, and this further complicates the decoding procedure.
	
	Nevertheless, after proper deformation of the initial graph, redefinition of weights of edges, and some changes in the procedure of finding detection events, 
	it is still possible to reduce the task of decoding to the Blossom algorithm for finding minimum-weight perfect matching. 
	Below we provide the description of modifications and motivation for them using a particular example.
	
	Consider a system with 3 data qubits $q_1,q_2,q_3$ and 1 ancilla qubit $a$. 
	We note that the ancilla qubit is moving from one place to another relative to data qubits. 
	This movement is due to the execution of riffle elements described in Fig.~\ref{Schemes_3q} (a) and (b). 
	Suppose that one bit-flip error occurred at time step $t = 2$ in the data qubit $q_1$ and another bit-flip error occurred at step 3 in the ancilla qubit $a$. 
	Error is denoted by a red (dark gray) thick edge. Blue (light gray) nodes correspond to measurements of a syndrome. 
	Since on the step $t = 2$ we perform the only measurement defined by the operator $ZIZ$, but not by the operator $ZZI$, we cannot tell if an error in data qubit $q_3$ or in data qubit $q_1$ is the most probable event. 
	The same is true for the step $t = 3$. 
	The first time step after $t = 2$ for which a measurement of $ZZI$ is performed is $t = 4$. 
	Thus, there is a need for connection of blue (light gray) node at $t = 2$ and blue (light gray) node at $t = 4$. 
	This connection will correspond to the error in the qubit $q_1$ at time step $t = 2$, if nodes of this connection will be marked as detection events and connected in the final matching.
	
	Now suppose a bit-flip error occurred at time step $t = 3$ in the ancilla qubit $a$. 
	At time $t = 3$ detection event may correspond to either error in $q_3$, or $q_1$, or $a$. before we make further measurements, we are not able to tell which error is the most probable. 
	Suppose that measurements during steps $t = 4$ and $t = 5$ did not detect difference between qubits $q_1$ and $q_2$, and measurement at the step $t = 6$ did not detect difference between qubits $q_3$ and $q_1$. 
	In this case, the error in the qubit $a$ on the step $t = 3$ is the most probable event, since any other set of errors that can produce the same syndrome would contain more than one error, and thus would be less probable than single error. 
	In order to have the ability to make such conclusions with procedure operating with the graph, 
	we should have the connection between the blue (light gray) node at step $t = 3$ and the blue (light gray) node at step $t = 6$, which would correspond to the error in $a$ at the step $t = 3$. 
	Analyzing in a similar way every possible single-qubit error, we come up with the graph structure illustrated in Fig.~\ref{RepetitonGraphProblems_DecodingAlgorithm}(b).
	
	Let us follow all the calculations with the syndrome corresponding to the two mentioned errors during steps $t = 2$ and $t = 3$. 
	Now we have not two, but one measurement at any particular time, thus we would need only one variable $m^t$ indexed by time steps, not two.
	Let $m^t$ be equal to $x\in\{0,1\}$ if the state of $a$ at $t$ is measured to be $|x\rangle$. 
	Having syndrome $m^t$, we can calculate indicators of a ``flip'' of ancilla $\lambda^t = m^t\oplus m^{t-1}$. 
	Detection events $d^t$ are calculated a bit differently. 
	Detection event should correspond to a change in the pattern of ``flipping'' for measurements of a particular generator. 
	In case of simultaneous measurements we have a set of ``flipping'' indicators $\lambda^t_1$ for generator $ZIZ$, and a set of ``flipping'' indicators $\lambda^t_2$ for generator $ZZI$. 
	Thus, we were able to build detection event indicators independently for $ZIZ$ and $ZZI$ by simply applying XOR to the corresponding variables. 
	But in the case of consecutive measurements, variable $\lambda^t$ contains indicators of ``flipping'' for both $ZIZ$ and $ZZI$, depending on time step $t$. 
	Thus, to properly get detection events, we should apply XOR to variables $\lambda^t$ at time steps corresponding to the same generators. 
	For instance, $d^3 = \lambda^3\oplus\lambda^2$, because measurements defined by $ZIZ$ were performed during steps $t = 3$ and $t = 2$. $d^6 = \lambda^6\oplus\lambda^3$, 
	because measurements defined by $ZZI$ were performed during steps $t = 6$ and $t = 3$ and so on.
	
	But before applying the Blossom algorithm to the subgraph, we have to specify the weights of edges. 
	In the case of simultaneous measurements, bit-flip errors in both data- and ancilla-qubits were assumed to be independent and equal. 
	In the case of consecutive measurements, 
	we use the same probabilistic model, but edges in the deformed graph correspond now to different sets of events. 
	Consider, for example, the edge in Fig.~\ref{RepetitonGraphProblems_DecodingAlgorithm}(b) corresponding to the qubit $q_2$ at the moment $t = 4$ and the edge corresponding to the qubit $q_2$ at the moment $t = 5$. 
	If the edge for $t = 5$ indicates error, then this error in the qubit $q_2$ could have happened between $t = 4$ and $t = 5$, i.e. during 1 time step. 
	Let us denote the probability of such error as $p$. But if the edge for $t = 4$ indicates error, then this error could have happened between $t = 1$ and $t = 4$, i.e. during 3 time steps. 
	Assuming that errors are independent, the probability of error during 3 time steps will be $3p$. 
	This difference in the probabilities is indicated by the fact that the edge for qubit $q_2$ at time $t = 5$ is denoted by a single line, whereas for the $t = 4$ it is denoted by triple line.
	
	In defining edges' weights, we should keep in mind that the general task is to identify the most probable event from the subgraph. 
	Therefore weights of edges should be such that the minimization of edges' total weight in the matching corresponds to the maximization of the probability of the chosen error pattern. 
	In Appendix~\ref{sec:probabilities} we prove that in order to satisfy the desired condition, 
	the weight $w_i$ of the edge $i$ should be the following function of the probability of error $p_i$ indicated by this edge:
	\begin{equation}
		w_i = -\log(p_i).
	\end{equation}
	Having indicators for detection events, we can build a subgraph and apply the Blossom algorithm for this properly weighted subgraph. 
	In Fig.~\ref{RepetitonGraphProblems_DecodingAlgorithm}(c) one can see that in this particular example, 
	there are two possible minimum weight perfect matchings: both of them represent a reasonable set of errors for the observed syndrome. 
	The first matching properly decodes the true error process, and the second matching would be wrongly interpreted as an error in $a$ at the $t = 2$ and an error in $q_1$ at the $t = 4$. 
	Such misinterpretations arise from the fact that we use consecutive instead of similar measurements in a situation when two errors occurred in close time periods. 
	Thus, we have less information about the error process. 
	But for sufficiently sparse errors no misinterpretations would arise, because distant single-qubit errors would have unique matching.
	
	Note that the described algorithm is easily generalized for arbitrary odd number of qubits in the repetition code. 
	See Appendix~\ref{sec:five-decoding} for an example of the decoding graph in a five-qubit repetition code.
	
	\subsection{Decoding in the Laflamme's 5-qubit code}
	
	For the case of the five-qubit code with the {generating set} $\{ ZXXZI,XXZIZ,XZIZX,ZIZXX\}$, due to the more complex structure of the generators we cannot represent measurements in the form of a graph in a straight-forward way, 
	i.e. by repeating spatial pattern of qubits, as for the repetition code. 
	Nevertheless, there is some structure that can be exploited for efficient decoding in sufficiently sparse error patterns. 
	To explore this structure, let us deviate from the space ordering of data qubits, and use the structure of generators to draw the graph for decoding.
	
\begin{center}
		\begin{figure*}[!ht]
			\includegraphics[scale=0.88]{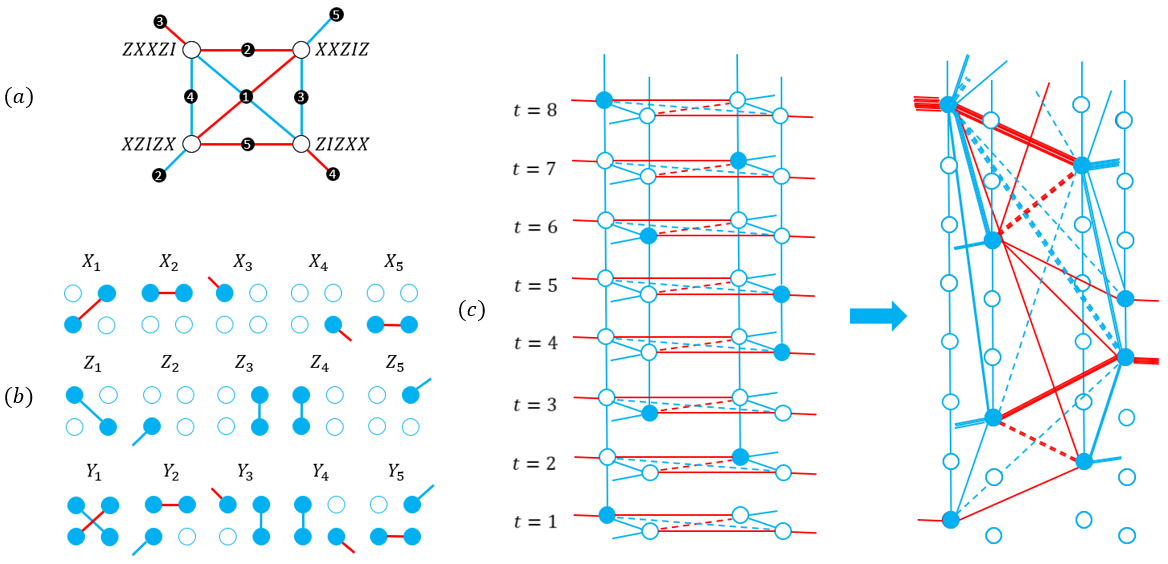}
			\caption{Details on the decoding procedure for the Laflamme's 5-qubit code. 
				(a) Graph representing the structure of the Laflamme's 5-qubit code's {generating set} at the fixed moment of time. (b) Error patterns for the 5-qubit code. 
				(c) Graph representing information from syndrome measurements (on the left-hand side) and deformed graph that takes into account consecutive nature of measurements and can be used for decoding.}
			\label{5QConnectivityErrorPatternsGraph}
		\end{figure*}
	\end{center}

	Let us depict ancilla qubits as empty nodes at vertices of a square. 
	Every ancilla qubit has a corresponding operator from the {generating set}, as shown in Fig.~\ref{5QConnectivityErrorPatternsGraph}(a).
	
	Now let us draw edges corresponding to $X$ and $Z$ entries in generators with the following rule: 
	for every data qubit with index $i$ we find a pair of the generators containing $X$ entry (or $Z$ entry) on the $i^{th}$ place and draw edge denoted by red (dark gray) color for $X$ 
	(or blue (light gray) color for $Z$) between the pair of ancilla nodes for corresponding generators. 
	For instance, consider data qubit with index $i = 1$. 
	We note that generators $XXZIZ$ and $XZIZX$ contain $X$ in the first place, and thus we draw a red (dark gray) edge between nodes for elements $XXZIZ$ and $XZIZX$ and denote this edge by the index $i = 1$. 
	
	Additionally, note that generators $ZXXZI$ and $ZIZXX$ contain $Z$ in the first place, and thus we draw a blue (light gray) edge between nodes for elements $ZXXZI$ and $ZIZXX$ and denote this edge by the index $i = 1$. 
	All edges are shown in Fig.~\ref{5QConnectivityErrorPatternsGraph}(a).
	
	These edges have the following meaning: if $Z$-error occurred in the data qubit with index $i$, then nodes adjacent to the red (dark gray) edge with index $i$ will give information about the change in the data qubit. 
	Correspondingly, if $X$-error occurred in data qubit with index $i$, then nodes adjacent to the blue (light gray) edge with index $i$ will give information about the change in the data qubit. 
	Finally, $Y$-error in data qubit with index $i$ can be identified by changes in measurement patterns of nodes adjacent to both red (dark gray) and blue (light gray) edges with index $i$. 
	Patterns for all single-qubit Pauli errors are shown in Fig.~\ref{5QConnectivityErrorPatternsGraph}(b). 
	Note that $Y$-errors produce syndromes for which 3 or even 4 nodes correspond to a single ``connection'' composed of two edges with different colors.
	
	As in the case of repetition code, we can construct a graph corresponding to consecutive measurements using linear connectivity and a single ancilla for the Laflamme's 5-qubit code for time steps $t$. 
	Making proper deformations to account for the consecutive nature of measurements, we get the graph which is shown on the right-hand side in Fig.~\ref{5QConnectivityErrorPatternsGraph}(c).
	
	When a syndrome is obtained, we can calculate detection events in the same manner as we did in the case of repetition code for consecutive measurements. 
	Using information about detection events, we can build a subgraph. 
	An important difference is the presence of $Y$-error, because the straight-forward implementation of the Blossom algorithm to find minimum weight perfect matching in the subgraph would not work properly due to the structure of weights. 
	Suppose, for simplicity, that all edges have equal weights, and we obtain a subgraph on ancillas corresponding to elements $ZXXZI$, $XXZIZ$ and $ZIZXX$. 
	Such a syndrome corresponds to the error $Y_3$ with matching consisting of two edges: red (dark gray) one and blue (light gray) one. 
	But another matching with the same weight is possible: consider a scenario when two errors $X_2$ and $X_4$ occurred simultaneously. 
	This matching has two red (dark gray) edges, and thus the same weight, but the probability of two simultaneous errors $X_2$ and $X_4$ is significantly lower than the probability of single error $Y_3$. 
	In this example, the Blossom algorithm fails to properly detect errors due to the fact that $Y$ errors should be regarded as having 1 connection for 3 nodes 
	(or 1 connection for 4 nodes in case of the error $Y_1$) with the weight of a single error for this multi-node connection. 
	Thus, we have the task of finding a minimum weight perfect matching not on a graph, but on a hypergraph.
	
	At the moment of writing this paper, the authors were not able to find an efficient solution to the general task of finding a minimum weight perfect matching on a hypergraph. 
	And therefore we developed the following decoding procedure that is efficient for sufficiently sparse single-qubit errors: after the subgraph is defined by calculating detection events, we apply three steps. 
	In the first step, we search for all sets of 4 nodes which can correspond to the error $Y_1$.
	All such sets are interpreted as $Y_1$ errors, and corresponding nodes are excluded from the subgraph. In the second step, we search for all sets of 3 nodes which can correspond to errors $Y_2, Y_3, Y_4$, or $Y_5$. 
	Then, again, such sets are excluded from the graph and interpreted as corresponding errors. And finally, when there is no connection in the graph with 3 or 4 nodes, we can apply the Blossom algorithm to decode errors of $X$- and $Z$-type. 
	Such a procedure is efficient because all 3 steps can be processed efficiently. 
	However, the main requirement for the procedure to work is sufficient sparsity of errors. 
	In the case when many errors occur in a short period of time, the described procedure could fail to find the most probable set of errors.
	
	{Lastly, it is worth mentioning that the current version of decoder can be improved by taking into account the propagation of errors through gates, but incorporating of this modification in the sequential scheme is subject to further research.}
	
	\section{Simulations}\label{sec:simulations}
	
	To analyze error levels for which given circuit provides improvements in the fidelity of encoded state, we provide simulations of our schemes. 
	To simulate error process we use the following set of noise channels:
	\begin{itemize}
		\item \textbf{Amplitude damping channel} with Kraus operators of the form:
		\begin{equation}
			K^{\text{AD}}_1 = \begin{pmatrix}
				1 & 0\\
				0 & \sqrt{1-\gamma_\text{a}}\\
			\end{pmatrix},
		\end{equation}
		\begin{equation}
			K^{\text{AD}}_2 = \begin{pmatrix}
				0 & \sqrt{\gamma_\text{a}} \\
				0 & 0
			\end{pmatrix},
		\end{equation}
		where $\gamma_\text{a} = 1-e^{-\frac{T_g}{T_1}}$, $T_g$ is the duration of a gate, and $T_1$ is the relaxation time.
		\item \textbf{Phase damping channel} with Kraus operators that are as follows:
		\begin{equation}
			K^{\text{PD}}_1 = \begin{pmatrix}
				1 & 0\\
				0 & \sqrt{1-\gamma_\text{p}}\\
			\end{pmatrix}
		\end{equation}
		\begin{equation}
			K^{\text{PD}}_2 = \begin{pmatrix}
				0 & 0 \\
				0 & \sqrt{\gamma_\text{p}}
			\end{pmatrix},
		\end{equation}
		where $\gamma_\text{p} = 1-e^{-2T_g(\frac{1}{T_2}-\frac{1}{2T_1}})$ and $T_2$ is the dephasing time. 
		
		Total action of amplitude damping and phase damping channels can be represented by the quantum operation
		\begin{equation}
			D^{\text{AP}}_{T_g}(\rho) = \sum_i K^{\text{PD}}_i(\sum_j K^{\text{AD}}_j(\rho)K^{\text{AD}\dagger}_{j})K^{\text{PD}\dagger}_{i}.
		\end{equation}
		\item 
		\textbf{Depolarizing channel for one-qubit gate} with the following Kraus operators:
		\begin{equation}
			K^{\text{D}_1}_1 = \sqrt{1-\frac{3}{4}p_1}\sigma_0,\;K^{\text{D}_1}_2 = \sqrt{\frac{p_1}{4}}\sigma_1,
		\end{equation}
		\begin{equation}
			K^{\text{D}_1}_3 = \sqrt{\frac{p_1}{4}}\sigma_2,\;K^{\text{D}_1}_2 = \sqrt{\frac{p_1}{4}}\sigma_3,
		\end{equation}
		where $p_1$ is the parameter corresponding to the probability of depolarization and $\{\sigma_j\}_1^3$ is a set of standard Pauli matrices $\sigma_1 = X,\;\sigma_2 = Y,\;\sigma_3 = Z,$ and $\sigma_0$ denotes $I$.
		Total action of the channel can be represented by the quantum operation
		\begin{equation}
			D^{\text{D}_1}(\rho) = \sum_i K^{\text{D}_1}_i(\rho)K^{\text{D}_1\dagger}_{i}.
		\end{equation}
		\item 
		\textbf{Depolarizing channel for two-qubit gate} with the following Kraus operators:
		\begin{equation}
			K^{\text{D}_2}_{ij} = \begin{cases}
				\sqrt{1-\frac{15}{16}p_2}\sigma_i\otimes\sigma_j, &\text{if $i = j = 0$}\\
				\sqrt{\frac{1}{16}p_2}\sigma_i\otimes\sigma_j, &\text{otherwise}
			\end{cases},
		\end{equation}
		where $p_1$ is a parameter corresponding to the probability of depolarization
		\begin{equation}
			D^{\text{D}_2}(\rho) = \sum_{i,j} K^{\text{D}_2}_{ij}(\rho)K^{\text{D}_2\dagger}_{ij}.
		\end{equation}
		\item \textbf{Measurement error} with the following Kraus operators:
		\begin{equation}
			K^{\text{M}}_0 = \sqrt{1-p_m}\sigma_0,
		\end{equation}
		\begin{equation}
			K^{\text{M}}_1 = \sqrt{p_m}\sigma_1,
		\end{equation}
		where $p_m$ is the parameter corresponding to the probability of measurement error
		\begin{equation}
			D^{\text{M}}(\rho) = \sum_i K^{\text{M}}_i(\rho)K^{\text{M}\dagger}_{i}.
		\end{equation}
	\end{itemize}
	
\begin{center}
		\begin{figure}[!ht]
			\includegraphics[scale=0.9]{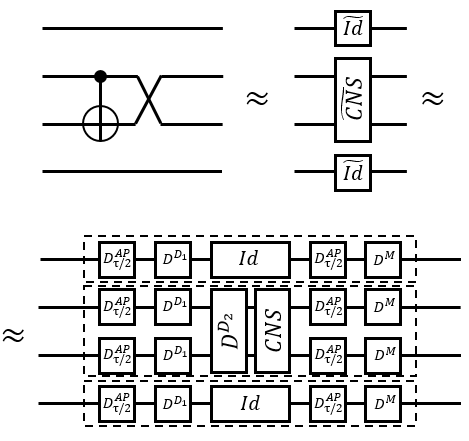}
			\caption{Example of the noise channel for CNS gate applied to two out of four qubits.}
			\label{Noise_channels}
		\end{figure}
	\end{center}

	To properly simulate behavior of the circuit, we must not only associate noise channels with one- and two-qubit gates, but also simulate noise in qubits which are left untouched during execution of other gates. 
	For this purpose, we use identity gates $Id$ and add noise to those gates. 
	Examples of noise channels associated with one- and two-qubit gates are depicted in Fig.~\ref{Noise_channels} and can be written in the following way:
	\begin{equation}
		\begin{split}
			& \Tilde{Id} = D^{\text{M}}\cdot D^{\text{AP}}_{\tau/2}\cdot D^{\text{D}_1}\cdot D^{\text{AP}}_{\tau/2}, \\
			& \Tilde{H} = D^{\text{M}}\cdot D^{\text{AP}}_{\tau/2}\cdot H\cdot D^{\text{D}_1}\cdot D^{\text{AP}}_{\tau/2}, \\
			& \Tilde{CNS} = \big(D^{\text{M}}\otimes D^{\text{M}}\big)\cdot \big(D^{\text{AP}}_{\tau/2}\otimes D^{\text{AP}}_{\tau/2}\big)\cdot CNS\cdot \\
			& \cdot D^{\text{D}_2}\cdot \big(D^{\text{D}_1}\otimes D^{\text{D}_1}\big)\cdot \big(D^{\text{AP}}_{\tau/2}\otimes D^{\text{AP}}_{\tau/2}\big).
		\end{split}
	\end{equation}
	
	In our simulation, we assume that $\tau = 13\cdot T_g$, since time $\tau$ of execution of 1 CNOT gate is approximately 13 times greater than time $T_g$ of execution of a single gate 
	(as estimated for IBM Belem quantum processor between qubits 1 and 2). 
	It serves as the worst-case benchmark scenario for the time of execution of CNS gate.
	
	\subsection{3-qubit code}
	
	In the 3-qubit code, we encode logical $|1\rangle$ state into physical $|111\rangle$ state of three data qubits. 
	At the end of the circuit, we take a snapshot of a density matrix $\rho$ of three data qubits and then find correction transformation $U$ according to the procedure for consecutive measurements described in Sec.~\ref{sec:decoding}. 
	The density matrix of the corrected state is $\rho_c = U\rho U^\dagger$. 
	Then we calculate fidelities for corrected and uncorrected states using reduced formulas:
	\begin{equation}
		\begin{split}
			F(|111\rangle\langle111|,\rho) = \sqrt{\langle111|\rho|111\rangle}, \\
			F(|111\rangle\langle111|,\rho_c) = \sqrt{\langle111|\rho_c|111\rangle}.
		\end{split}
	\end{equation}
	
\begin{center}
		\begin{figure}[!ht]
			\includegraphics[scale=0.43]{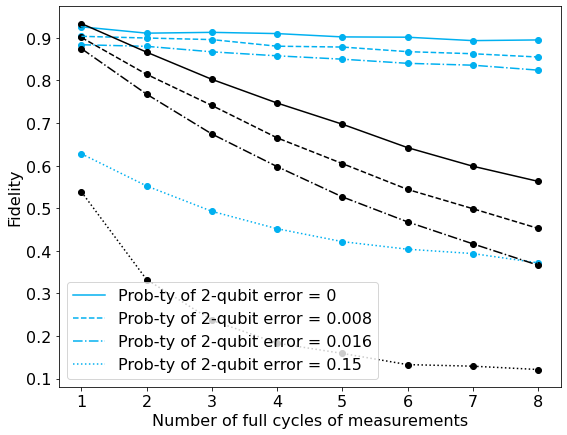}
			\caption{Fidelities of uncorrected (black lines) and corrected (blue (light gray) lines) states for different levels of two-qubit error probabilities as functions of cycles of measurements in the 3-qubit repetition code.}
			\label{Simulations3_1}
		\end{figure}
	\end{center}

	Results of scheme simulations for different levels of a two-qubit depolarizing error are presented in Fig.~\ref{Simulations3_1}.
	Y-axis corresponds to the fidelity and X-axis corresponds to the number of cycles. 
	Here 1 cycle is equivalent to 2 series of full {generating set} measurements, totaling 4 measurements per 1 cycle. 
	
	Every line style (solid, dashed, dotted or dot-and-dash) corresponds to some particular probability of errors in two-qubit gates. 
	For every line style, there are 2 colors: the black color representing fidelities for uncorrected states and the blue (light gray) color representing fidelities for corrected states. 
	For every line on the graph, error parameters for the noise process are equivalent except for a two-qubit depolarizing error. 
	Characteristics of a noise process are set equal to the measured corresponding parameters of the qubit with index 0 from the IBM Belem quantum processor. 
	Those characteristics are given in Appendix~\ref{sec:characteristics}.
	
	It can be seen from Fig.~\ref{Simulations3_1} that for every chosen level of two-qubit error, implementation of the scheme provides significant improvement in the fidelity of the preserved state.
	
	\begin{center}
		\begin{figure}[!ht]
			\includegraphics[scale=0.43]{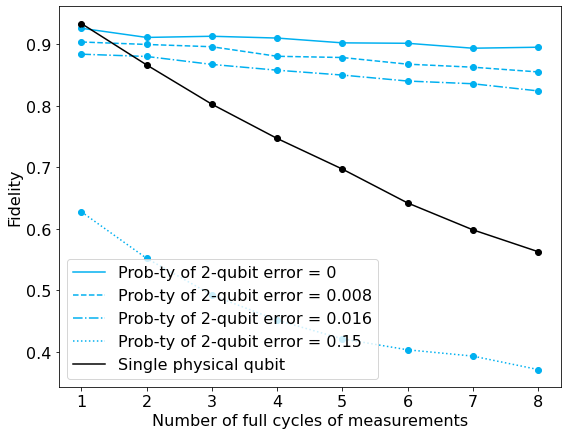}
			\caption{Fidelities of corrected (blue (light gray) lines) states for different levels of two-qubit error levels in comparison with fidelities of preservation of the state $|1\rangle$ in 1 physical qubit (black line).}
			\label{Simulations3_2}
		\end{figure}
	\end{center}
	
	For the benchmark, we can use the single-qubit state initialized as $|1\rangle$ and preserved under single-qubit noise for the time of execution of $k$ cycles of the scheme and 
	compare the resulting fidelity for single-qubit preservation of state $|1\rangle$ with fidelity for the preservation of the state $|111\rangle$ with repetition code using correction procedures. 
	Results are presented in Fig.~\ref{Simulations3_2}.
	
	One can see that the fidelity of a logical corrected state preserved by repetition code for sufficiently low two-qubit errors is higher than the fidelity of a state in a single qubit without applying repetition code procedures.
	
	{Additionally, if we compare the black solid line in Fig.~\ref{Simulations3_2} to our experimental results in Fig.~\ref{IBM_Experiments_New} (blue (light gray) solid line), we can see that fidelities of logical qubits encoded in IBM Belem, IBM Quito and IBM Manila are close to or exceed the fidelity calculated for single physical qubit. This suggests that our scheme can help to reach the break-even point for repetition code where the logical qubit has longer coherence time than that of the constituting physical qubits.}
	
	\subsection{Laflamme's 5-qubit code code}
	
	In simulations for the five-qubit code, the initial state is as follows:
	\begin{center}
		$|\psi_0\rangle = \frac{1}{4}[1,0,0,1,0,-1,1,0,$
		
		$\quad \quad \quad \quad \quad \quad 0,-1,-1,0,1,0,0,-1,$
		
		$\quad \quad \quad \quad \quad \quad 0,1,-1,0,-1,0,0,-1,$
		
		$\quad \quad \quad \quad \quad \quad 1,0,0,-1,0,-1,-1,0].$
	\end{center}
	
	It can be easily checked that the state $|\psi_0\rangle$ is stabilized by every element of the {generating set} for the Laflamme's 5-qubit code, and thus $|\psi_0\rangle$ is in the code space. 
	
\begin{center}
		\begin{figure}[!ht]
			\includegraphics[scale=0.43]{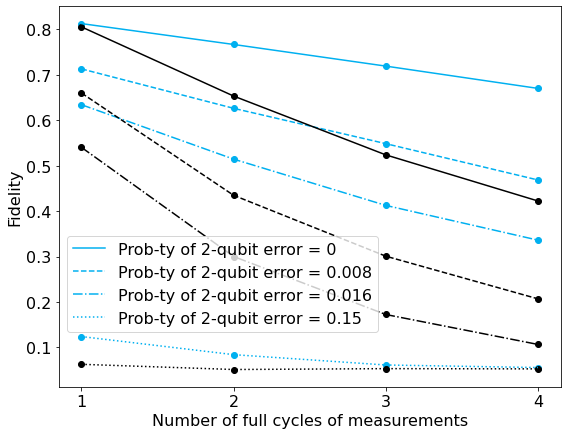}
			\caption{Fidelities of uncorrected (black lines) and corrected (blue (light gray) lines) states for different levels of two-qubit error probabilities as functions of cycles of measurements in the Laflamme's 5-qubit code.}
			\label{Simulations5}
		\end{figure}
	\end{center}

	In the graph from Fig.~\ref{Simulations5} all lines represent fidelities of a logical {qubit} encoded into 5 physical data qubits. 
	The setting is quite similar to the 3-qubit case, but we provide detailed description to avoid misunderstanding. 
	Y-axis corresponds to the fidelity, X-axis corresponds to the number of measurement cycles, and every line style (solid, dashed, dotted or dot-and-dash) corresponds to some particular probability of errors in two-qubit gates. 
	Here 1 cycle is equivalent to 2 series of full {generating set} measurements, totalling 8 measurements per 1 cycle. 
	For every line style, there are 2 colors: the black color representing fidelities for uncorrected states and the blue (light gray) color representing fidelities for corrected states. 
	For every line on the graph, error parameters for the noise process are equivalent except for a two-qubit depolarizing error. 
	Characteristics of a noise process are set equal to the measured corresponding parameters of the qubit with index 0 from the IBM Belem quantum processor. 
	Those characteristics are given in Appendix~\ref{sec:characteristics}.
	One can see that the correction procedure does increase the fidelity of a state, although high levels of a two-qubit error lead to insignificant improvements in the fidelity. {Thus, with decrease in probablility of two-qubit errors, our scheme could be helpful for achieving fidelities of a logical qubit that exceed fidelity of single physical qubit (compare black solid line in Fig.~\ref{Simulations3_2} with blue (light gray) lines in Fig.~\ref{Simulations5}). But at the moment of conducting the experiment, such low levels of probability of two-qubit errors were not available.}
	
	\section{Experiments using IBM quantum processors}\label{sec:IBM}
	
	At the moment of writing this paper, several IBM superconducting quantum processors with the ability to execute mid-circuit measurements were available. 
	The maximal amount of qubits for these processors is five, which does not allow to carry out experiments for the 5-qubit code, since one additional ancilla qubit is required. 
	An additional complication in using a quantum processor for experiments with the 5-qubit code would be the necessity for quantum tomography of the resulting state to get the fidelities.
	
	But we can carry out experiments with the 3-qubit repetition code, since 4 qubits are enough for such experiments, and there is no need for the tomography: 
	the fidelity between resulting state and the state $|111\rangle$ could be estimated as the square root of the fraction of measured state $|111\rangle$ after measuring the 3 data qubits in computational basis at the end of the circuit many times. 
	Correction procedures are represented by bit-flips of corresponding qubits, so the correction could be executed classically after the measurement of the final state.
	
\begin{center}
		\begin{figure}[!ht]
			\includegraphics[scale=0.7]{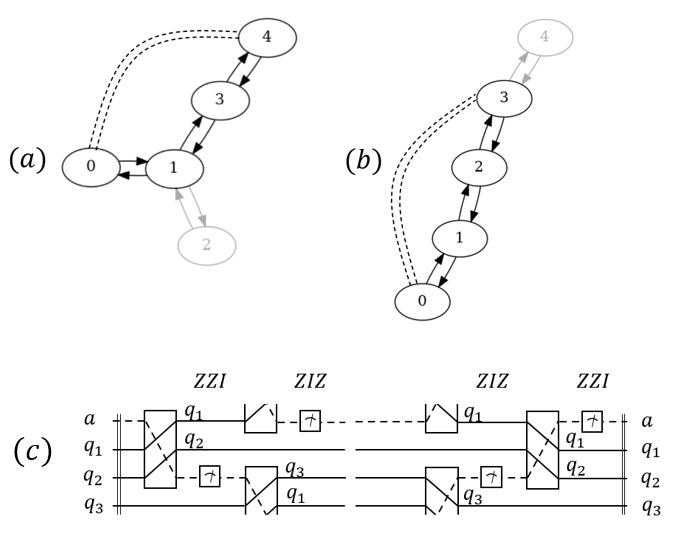}
			\caption{In (a) Y-type connectivity pattern of IBM quantum processors is illustrated.
				In (b) I-type connectivity pattern is shown.
				Light gray color denotes a qubit which is not used in our scheme.
				In (c) the circuit realizing 1 cycle of consecutive syndrome measurements in our scheme for the 3-qubit repetition code is demonstrated.}
			\label{4QSARepetitionScheme}
		\end{figure}
	\end{center}

	Among available quantum processors, there are two types of connectivity patterns depicted in Fig.~\ref{4QSARepetitionScheme} (a) and (b). 
	We will refer to them as ``Y-type'' and ``I-type'' connectivity patterns for (a) and (b) respectively. Nodes correspond to qubits and edges correspond to physical couplings of these qubits. 
	Dotted edges represent lacking physical connections necessary for circular pattern of connectivity. 
	Light gray color denotes a qubit which is not used in the realization of our scheme. 
	Other 4 qubits are used for realization of the scheme from the Fig.~\ref{4QSARepetitionScheme}(c).
	
	We decided to use 4 physical qubits for two reasons: firstly, it is important that the connectivity between qubits is as close to the circular pattern as possible. 
	Quantum processors of the Y-type in Fig.~\ref{4QSARepetitionScheme}(a) allow for linear pattern of connectivity with 4 qubits, but if we used 5 qubits instead, we would not have had an opportunity to exploit the simple connectivity pattern. 
	Secondly, when only 3 qubits are used as data qubits, we are able to make a comparison between our scheme with 4 physical qubits (3 data qubits and 1 ancilla qubit) and the scheme with 5 physical qubits (3 data qubits and 2 ancilla qubit) realizing the same repetition code. 
	
	\begin{center}
		\begin{figure}[!ht]
			\includegraphics[scale=0.35]{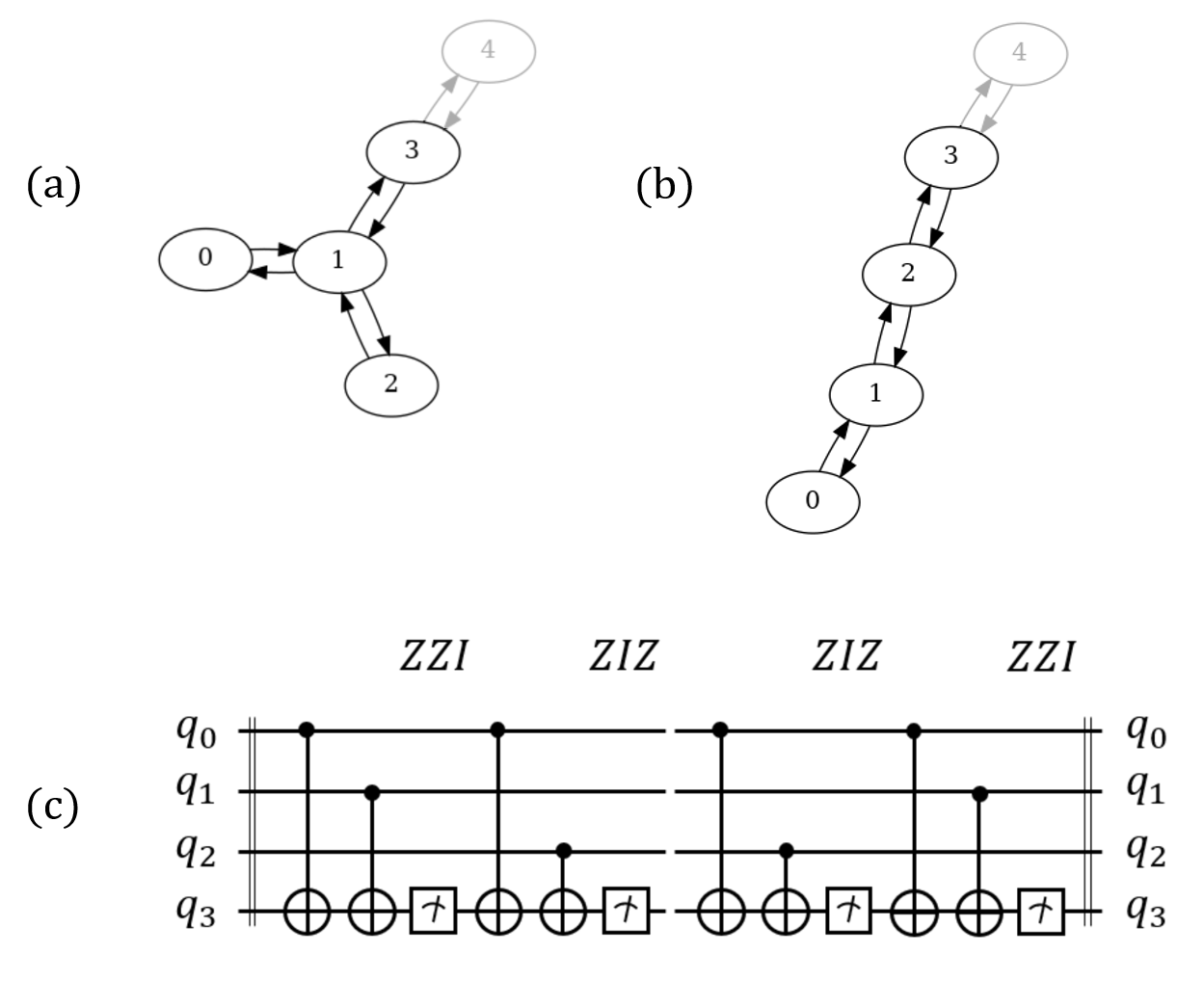}
			\caption{In (a) Y-type connectivity pattern of IBM quantum processors is illustrated.
				In (b) I-type connectivity pattern is shown.
				Light gray color denotes a qubit which is not used in the benchmark scheme.
				In (c) the circuit realizing 1 cycle of consecutive syndrome measurements in the benchmark scheme for the 3-qubit repetition code with 3 data qubits and 1 ancilla qubit is presented.}
			\label{4QRepetitionScheme}
		\end{figure}
	\end{center}
	
	\vspace{-0.6in}
	The list of available superconducting quantum processors with corresponding types of connectivity is the following:
	\begin{itemize}
		\item IBM Lima has Y-type;
		\item IBM Belem has Y-type;
		\item IBM Quito has Y-type;
		\item IBM Bogota has I-type;
		\item IBM Santiago has I-type;
		\item IBM Manila has I-type.
	\end{itemize}
	
	\begin{center}
		\begin{figure*}[!ht]
			\includegraphics[scale=0.7]{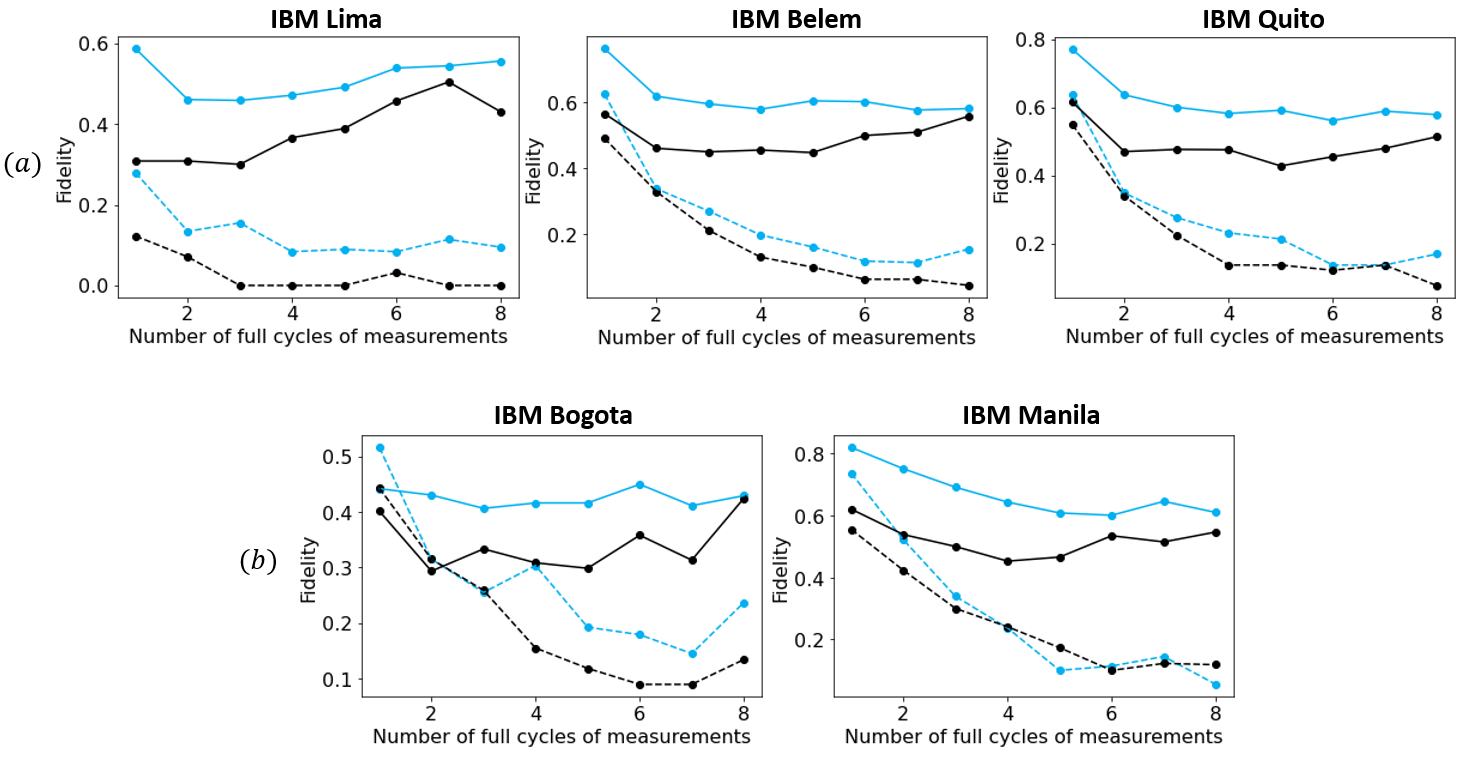}
			\caption{Comparison of fidelities in experiments on quantum processors. 
				Our scheme of repetition code using 4 physical qubits is compared with the well-known scheme of repetition code using 4 qubits. 
				Each point on the graph corresponds to an estimated fidelity of a state in 1 experiment with a given number of full cycles of measurements. 
				Blue (light gray) lines represent results of experiments with the realization of our scheme and black lines represent results for the well-known realization of the same repetition code. In both schemes, 4 physical qubits (3 data qubits and 1 ancilla qubit) are used. 
				Dotted lines depict fidelities of resulting states without correction. Experiments in (a) correspond to quantum processors with Y-type connectivity; experiments in (b) correspond to quantum processors with I-type connectivity.}
			\label{IBM_Experiments_New}
		\end{figure*}
	\end{center}
	
	\begin{center}
		\begin{figure}[!ht]
			\includegraphics[scale=0.6]{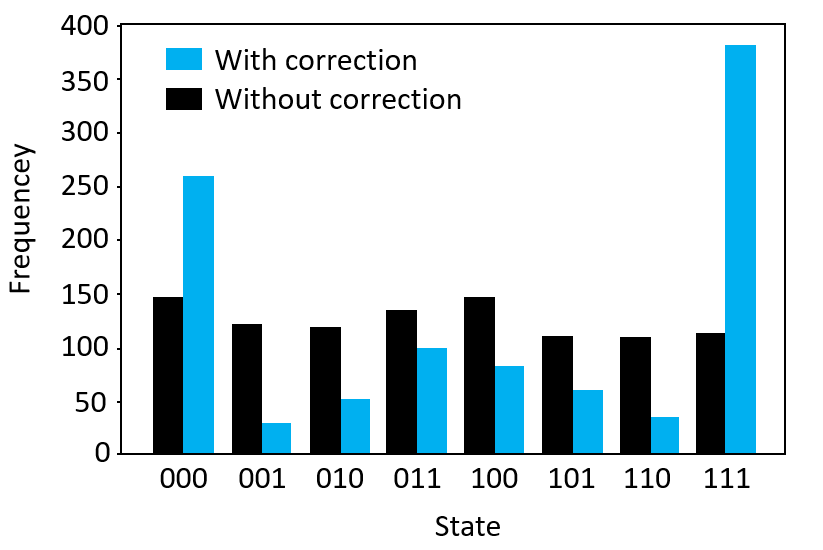}
			\caption{Frequencies of measured states from the experiment with IBM Belem quantum processor for 2 cycles of measurements using 1000 shots.}
			\label{IBM_Experiment2cycles}
		\end{figure}
	\end{center}
	
	\vspace{-0.8in}
	We note, however, that the architecture of all those processors has two features that reduce the potential effect of an increase in fidelity levels from application of the developed scheme. 
	The first feature is that processors lack 1 physical connection between qubits, and thus do not have a circular pattern of connectivity. 
	This lacking connection is denoted by the dashed line in Fig.~\ref{4QSARepetitionScheme} (a) and (b). 
	Presumably, this connection is transpiled into a sequence of SWAP gates via existing physical connections in the rest of qubits. 
	Many SWAP gates for the realization of the lacking connection could significantly increase levels of noise. 
	The second feature is the fact that we don't know if the operation CNOT$\cdot$SWAP is indeed realized as more efficient CNS with single-qubit gates rather than two consecutive CNOT- and SWAP-gates.
	
	Here we analyze a performance of our scheme using 4 physical qubits that is shown in Fig.~\ref{4QSARepetitionScheme}(c) and compare it with the well-known repetition code scheme with the same amount of physical qubits given in Fig.~\ref{4QRepetitionScheme}(c).
	In both schemes, 3 physical qubits are used as data qubits and 1 physical qubit is used as an ancilla qubit. 
	In the benchmark scheme, CNS operations are not represented and CNOT operations between the ancilla qubit and all other data qubits are exploited.
	This leads to a necessity of applying additional SWAP gates in the absence of full connectivity.
	In the benchmark scheme, one cycle contains four generators' measurements to match one cycle of measurements in our scheme.
	
	Results of experiments are shown in Fig.~\ref{IBM_Experiments_New}. 
	Quantum processor IBM Santiago was not available at the moment of conducting the experiment. 
	It is important to note that fidelity levels before the error-correction are quite close in both schemes. 
	Since after the correction our scheme shows significantly higher fidelity levels across all quantum processors and for any number of cycles, we conclude that our scheme performs better than the benchmark scheme. 
	
	If we compare these results with the results of simulation in Fig.~\ref{Simulations3_1}, we see that the physical system behaves similarly to the simulated system with a two-qubit error level of $0.15$. 
	It leads us to the assumption that a potential decrease in this error level by making use of CNS gates can significantly improve the system's performance.
	
	Graph representing frequencies of measured states for 1000 shots in the 3-qubit code scheme with 2 full cycles of measurements is shown in Fig.~\ref{IBM_Experiment2cycles}. 
	Experiment was carried out on IBM Belem. 
	One can see from the graph that correction leads to a significant increase in the frequency of measured states $|111\rangle$ and $|000\rangle$, 
	whereas other states have significantly reduced frequencies. 
	We make a conclusion that the scheme works well and correctly identifies a majority of bit-flip errors when errors are not concentrated densely around some particular period of time. 
	The incorrect decoding of the errors which result in decoded $|000\rangle$ state presumably corresponds to cases of a dense concentration of errors during some time-periods.
	
	Thus, our schemes provide better results than the state-of-the-art methods that use the same amount of required resources. 
	
	{In general, if we compare our scheme with schemes that use more ancilla-qubits, those schemes can have an advantage by allowing for parallel measurements. In the worst case, number of parallel measurements will be equal to number of ancilla-qubits and time for execution of one- and two-qubit gates in sequential and parallel schemes will be close. Then time $t_s$ required for performing full cycle of measurements in sequential scheme and time $t_p$ for parallel scheme will roughly relate as $t_s = nt_p$, where $n$ is number of ancilla-qubits in the parallel scheme. This means that although our scheme provides advantage in the space by exploiting less qubits, it requires more time-resources. Comparing sequential and parallel schemes by logical error rates for realistic noise models analytically is a challenging problem, so we provided experimental comparison for the case of our scheme using 4 physical qubits (3 data qubits and 1 ancilla qubit) in the repetition code and the well-known scheme using 5 physical qubits (3 data qubits and 2 ancilla qubits).} Results are provided in Appendix~\ref{sec:two-ancillas-experiments}. We note that performance of our scheme is close to the performance of the method that uses more space resources.
	
	{Another potential advantage of our scheme is that the exchange of data and ancillary qubits may help dealing with leakage errors, especially troublesome in superconducting qubits~\cite{Ghosh_2015}. But testing applicability of this idea is beyond the scope of the current paper.}
	
	\section{Conclusion}\label{sec:conlcusion}
	
	In this work, we have developed the scheme for the realization of the wide class of error correction codes using single ancilla and circular connectivity pattern.  
	Examples of codes that can be realized in this way include repetition codes, the Laflamme's 5-qubit code, Shor's 9-qubit code, and others.
	{For near-neighbour connectivity, we demonstrated how to apply our scheme to realize surface code with single ancilla.}
	Requirements for the implementation of the proposed scheme make it attractive for the superconducting quantum computing platform, 
	because our method has been developed to use CNS gates.
	We have conducted simulations of the circuits for 3-qubit repetition and Laflamme's 5-qubit codes, whose results demonstrate that fidelities of preserved state increase for various levels of two-qubit error and various numbers of cycles in circuits. 
	Moreover, our simulations have demonstrated that for sufficiently low levels of two-qubit errors significant increase in fidelities could be achieved by applying the suggested circuits.
	Fidelity of the logical {qubit} encoded into three data qubits after corrections could even exceed the fidelity of the state $|1\rangle$ preserved in 1 physical qubit. 
	Finally, experiments using IBM superconducting quantum processors were carried out to assess the performance of the suggested scheme for the 3-qubit repetition with only 4 physical qubits 
	(where 3 qubits were used as data qubits and 1 ancilla qubit was used for measurements). 
	Although gate combinations encoded as CNOT$\cdot$SWAP might not have been transpiled to low-noise CNS-gates and 1 physical connection to guarantee circular pattern of connectivity lacked, 
	fidelities of preserved state have been significantly improved. 
	Thus, we expect that our scheme is of interest for demonstrating error-correction codes with superconducting quantum processors.
	
	\vspace{-0.3in}
	\section*{Acknowledgements}
	We acknowledge use of the IBM Q Experience for this work. 
	The views expressed are those of the authors and do not reflect the official policy or position of IBM or the IBM Q Experience team.
	This work is supported by the Russian Roadmap on Quantum Computing (development of the error correction code in Sec.~\ref{sec:single-ancilla}, simulations in Sec.~\ref{sec:simulations}, and experiments in Sec.~\ref{sec:IBM}),
	by the grant of the Russian Science Foundation No. 19-71-10091 (generalization of the proposed method in Sec.~\ref{sec:generalization}),
	and by the Priority 2030 program at the National University of Science and Technology ``MISIS'' (realization of the efficient decoding procedure in Sec.~\ref{sec:decoding}).
	
	\appendix
	
	\section{Circuits}\label{sec:App1}
	
	Let us first define a projective measurement by observable $M$. 
	Consider spectral decomposition of Hermitian operator $M = \sum_{\{m\}} m\Pi_m$, where $\{m\}$ is a set of eigenvalues of $M$ and $\{\Pi_m\}$ is a set of projectors onto the corresponding eigenspaces of $M$.
	Performing a projective measurement defined by $M$ corresponds to the following:
	\begin{itemize}
		\item getting some particular outcome $m'$ from the set $\{m\}$ of eigenvalues of $M$, and
		\item projecting the state $|\psi\rangle$ to the state 
		\begin{equation}
			\frac{\Pi_{m'}|\psi\rangle}{\sqrt{\langle\psi|\Pi_{m'}|\psi\rangle}}.
		\end{equation}
	\end{itemize}
	The probability of the outcome $m'$ is defined as $p(m') = \langle\psi|\Pi_{m'}|\psi\rangle$.
	
\begin{center}
		\begin{figure}[]
			\includegraphics[scale=0.5]{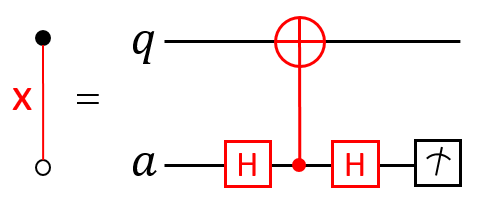}
			\caption{Circuit realizing measurement defined by $X$.}
			\label{Measurement_X}
		\end{figure}
	\end{center}

	Let us then suppose that $M = X$. Pauli matrix $X$ has two eigenvalues: $+1$ and $-1$. Its spectral decomposition is as follows:
	\begin{equation}
		\begin{split}
			X &= (1)\cdot\Pi^X_+ + (-1)\cdot\Pi^X_- \\
			&= (1)\cdot\frac{I+X}{2} + (-1)\cdot\frac{I-X}{2},
		\end{split}
	\end{equation}
	
	where $\Pi^X_+$ ($\Pi^X_-$) is a projector corresponding to $+1$ ($-1$) eigenvalue.
	Thus, only two outcomes corresponding to eigenvalues $+1$ and $-1$ are possible. 
	Let us show that the circuit in Fig.~\ref{Measurement_X} realizes the measurement defined by $X$.
	
	Suppose that a data qubit $q$ is in the state $|\psi\rangle$ and an ancilla qubit $a$ in the state $|0\rangle$. 
	Then before a measurement of the ancilla qubit at the end of the circuit is performed, the state of the composite two-qubit system takes the form
	
	\begin{center}
		\begin{figure}[]
			\includegraphics[scale=0.5]{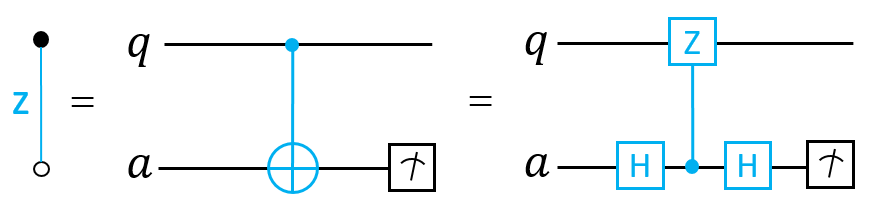}
			\caption{Circuit realizing measurement defined by $Z$.}
			\label{Measurement_Z}
		\end{figure}
	\end{center}
	
	\begin{center}
		\begin{figure*}[!ht]
			\includegraphics[scale=0.95]{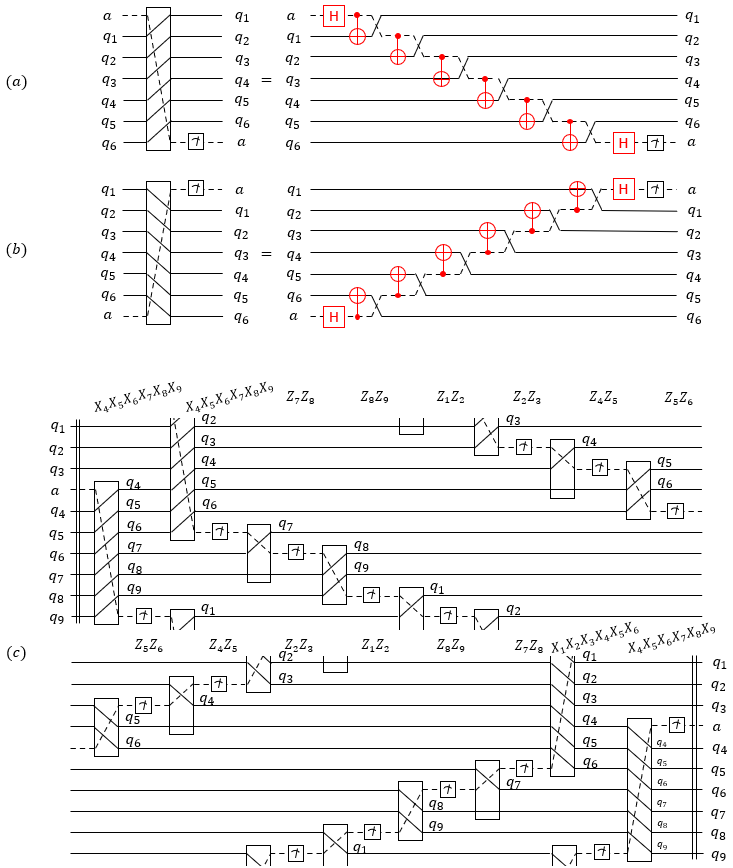}
			\caption{Realization of the nine-qubit code. 
				In (a) we show ``riffle'' block for a measurement defined by the operator $XXXXXXIII$. 
				In (b) we show ``reverse riffle'' block for a measurement defined by the operator $XXXXXXIII$. 
				In (c) the circuit realizing consecutive syndrome measurements for Shor's nine-qubit code is illustrated.}
			\label{Schemes_9q}
		\end{figure*}
	\end{center}
	
	\vspace{-1.2in}
	\begin{equation}
		\begin{split}
			&(I\otimes H)\cdot CNOT\cdot(I\otimes H)|\psi\rangle|0\rangle = \\
			&= (I\otimes H)\cdot CNOT\Big(\frac{|\psi\rangle|0\rangle+|\psi\rangle|1\rangle}{\sqrt{2}}\Big) \\ 
			& = (I\otimes H)\Big(\frac{|\psi\rangle|0\rangle}{\sqrt{2}}+\frac{X|\psi\rangle|1\rangle}{\sqrt{2}}\Big) = \\
			&= \frac{|\psi\rangle|0\rangle+|\psi\rangle|1\rangle}{2}+\frac{X|\psi\rangle|0\rangle-X|\psi\rangle|1\rangle}{2} = \\
			&= \frac{I+X}{2}|\psi\rangle|0\rangle+\frac{I-X}{2}|\psi\rangle|1\rangle = \\
			&= \Pi^X_+|\psi\rangle|0\rangle + \Pi^X_-|\psi\rangle|1\rangle.
		\end{split}
	\end{equation}
	
	Result a measurement for the ancilla qubit $a$ can be either $|0\rangle$ or $|1\rangle$. 
	If $|0\rangle$ is measured, then the state of the data qubit $q$ collapses to 
	\begin{equation}
		\frac{\Pi^X_+|\psi\rangle}{\sqrt{\langle\psi|\Pi^X_+|\psi\rangle}}.
	\end{equation} 
	If $|1\rangle$ is measured, then the state of the data qubit $q$ is as follows: 
	\begin{equation}
		\frac{\Pi^X_-|\psi\rangle}{\sqrt{\langle\psi|\Pi^X_-|\psi\rangle}}.
	\end{equation} 
	Thus, if we interpret the measurement of $|0\rangle$ in ancilla qubit as getting $+1$ eigenvalue and the measurement of $|1\rangle$ as getting $-1$ eigenvalue of observable $X$, 
	then action of the circuit on data qubit $q$ is equivalent to the projective measurement defined by the operator $X$.
	
	The fact that applying circuit from Fig.~\ref{Measurement_Z} is equivalent to performing a measurement defined by the operator $Z$ can be proven in the similar manner. 
	One just needs to notice that applying CNOT with control-qubit $q$ is equivalent to applying the composition of two Hadamard gates and controlled-$Z$ gate as it is shown in Fig.~\ref{Measurement_Z}.
	
	Suppose a data qubit $q$ is in the state $|\psi\rangle$ and an ancilla qubit $a$ in the state $|0\rangle$. 
	Then before a measurement of the ancilla qubit at the end of the circuit is performed, the state of the composite two-qubit system takes the following form:
	\begin{equation}
		\begin{split}
			&(I\otimes H)\cdot CZ\cdot(I\otimes H)|\psi\rangle|0\rangle = \\ 
			&= (I\otimes H)\cdot CZ\Big(\frac{|\psi\rangle|0\rangle+|\psi\rangle|1\rangle}{\sqrt{2}}\Big) \\
			&= (I\otimes H)\Big(\frac{|\psi\rangle|0\rangle}{\sqrt{2}}+\frac{Z|\psi\rangle|1\rangle}{\sqrt{2}}\Big) \\
			&= \frac{|\psi\rangle|0\rangle+|\psi\rangle|1\rangle}{2}+\frac{Z|\psi\rangle|0\rangle-Z|\psi\rangle|1\rangle}{2} \\
			&= \frac{I+Z}{2}|\psi\rangle|0\rangle+\frac{I-Z}{2}|\psi\rangle|1\rangle \\
			&= \Pi^Z_+|\psi\rangle|0\rangle + \Pi^Z_-|\psi\rangle|1\rangle.
		\end{split}
	\end{equation}
	Once again, if a measurement of the ancilla results in the state $|0\rangle$, it could be interpreted as an outcome corresponding to $+1$ eigenvalue. 
	Then the state $|1\rangle$, in turn, corresponds to eigenvalue $-1$.
	
	\section{Shor's 9-qubit code}\label{sec:nine}
	
	3- and 5-qubit schemes are of interest as resource-efficient realizations of 3- and 5-qubit codes, 
	whereas the nine-qubit code does not represent a significant deal of interest for realization in experimental settings: it possesses similar properties as the 5-qubit code and requires the use of more qubits and more gates. 
	Nevertheless, it is important to consider the proposed scheme for this famous error-correction code. 
	
\begin{center}
		\begin{figure}[!ht]
			\includegraphics[scale=0.33]{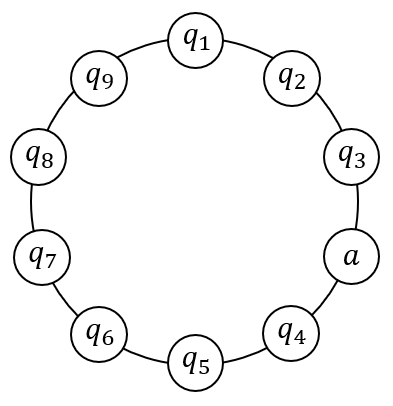}
			\caption{Circular connectivity pattern for 10 qubits.}
			\label{Chain9}
		\end{figure}
	\end{center}

	Suppose now we have 10 qubits connected in a circular chain as illustrated in Fig.~\ref{Chain9}: there are 9 data qubits $\{q_k\}_{k=1}^9$ ordered by their index number and 1 ancilla qubit.
	{Generating set} for the nine-qubit code includes 6 elements with $Z$-entries and 2 elements with $X$ entries. 
	To realize measurements defined by 
	\begin{equation}
		\begin{split}
			\{ZZIIIIIII,IZZIIIIII,IIIZZIIII, \\ IIIIZZIII,IIIIIIZZI,IIIIIIIZZ\}, 
		\end{split}
	\end{equation}
	we use gates that are shown in Fig.~\ref{Schemes_3q}(a) and Fig.~\ref{Schemes_3q}(b), 
	and for measurements defined by elements $\{XXXXXXIII,IIIXXXXXX\}$ we introduce more riffle gates as demonstrated in Fig.~\ref{Schemes_9q} (a) and (b). 
	The scheme realizing proper measurements is represented in Fig.~\ref{Schemes_9q}(c).
	
	\section{Modifications}\label{sec:Modifications}
	
\begin{center}
		\begin{figure}[]
			\includegraphics[scale=0.7]{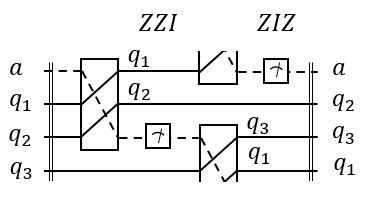}
			\caption{Modified circuit realizing 1 cycle of consecutive measurements defined by stabilizer elements $ZZI$ and $ZIZ$ in the 3-qubit repetition code.}
			\label{Circuit4Z_halved}
		\end{figure}
	\end{center}

	Here we describe several modifications of the scheme.
	For the first modification, we might have used the scheme from Fig.~\ref{Circuit4Z_halved} for repetitive measurements, 
	which is just the first part of the initial scheme.
	
	The scheme looks more compact, but there is an issue: the order of qubits after one cycle of measurements is permuted. 
	This means that if on the first step measurements defined by $ZZI$ and $ZIZ$ are executed, 
	then on the second step syndrome will be for elements $IZZ$ and $ZZI$. 
	However, in this particular case, {generating sets} $\{ ZZI,ZIZ\}$ and $\{ZZI,IZZ\}$ define the same stabilizer. 
	The same logic holds true for all the other cycles of measurements. 
	This implies that by means of classical post-processing we still can correct the same set of errors as for the initial code.
	
\begin{center}
		\begin{figure}[!ht]
			\includegraphics[scale=0.33]{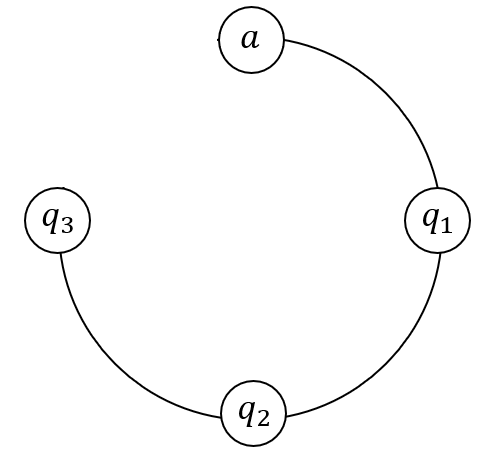}
			\caption{Reduced connectivity pattern for particular realization of the 3-qubit code.}
			\label{Chain3_reduced}
		\end{figure}
	\end{center}

\begin{center}
		\begin{figure*}[!ht]
			\includegraphics[scale=0.55]{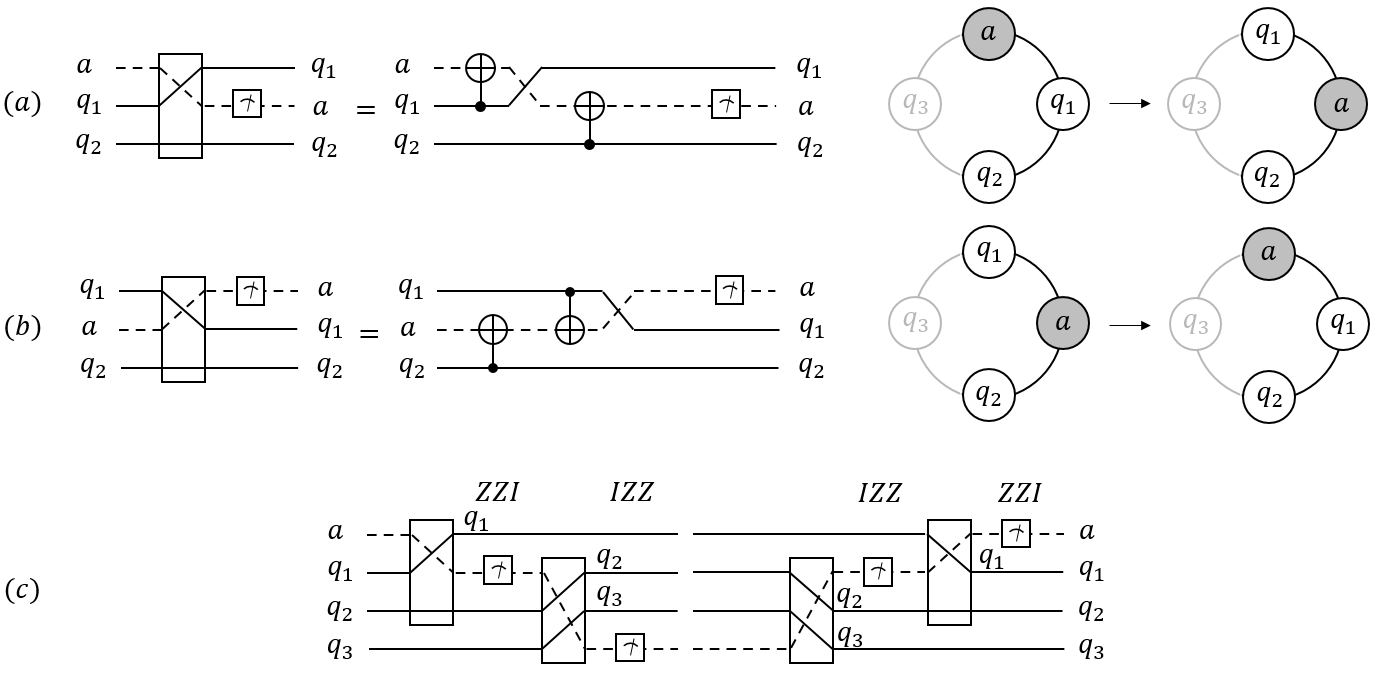}
			\caption{In (a) ``riffle with CNOT'' block for a measurement defined by the operator $ZZI$ is shown. 
				In (b) ``reverse riffle with CNOT'' block for a measurement defined by the operator $ZZI$ is illustrated. 
				In (c) the circuit realizing consecutive syndrome measurements for the 3-qubit repetition code with only 3 physical connections between 4 qubits is demonstrated.}
			\label{Schemes_3q_CNOT}
		\end{figure*}
	\end{center}

	The second modification may help us to get rid of one connection between qubits and use the connectivity pattern as illustrated in Fig.~\ref{Chain3_reduced}:
	However, the price for this improvement is the necessity to use CNOT-gates, which are not native for the superconducting platform. 
	We introduce two more types of riffle gates that contain CNS and CNOT in Fig.~\ref{Schemes_3q_CNOT} (a) and (b).
	
	Using all four types of riffle gates, we can build the circuit for repetitive syndrome measurements, as presented in Fig.~\ref{Schemes_3q_CNOT}(c). 
	In this case, syndrome is obtained from measurements defined by elements of the {generating set} $\{ZZI,IZZ\}$. 
	This {generating set} defines the same stabilizer as the initial {generating set} $\{ZZI,ZIZ\}$.
	
	\section{Probabilistic model for weights of edges}\label{sec:probabilities}
	
	Let us introduce some notations. 
	First of all, we prescribe index $i$ to every edge from the full graph. An example of the full graph in the case of the 3-qubit repetition code with simultaneous measurements using 5 physical qubits can be found in Fig.~\ref{RepetitonGraph_DecodingAlgorithm}(b).
	
	Consider the probabilistic model with the following properties:
	(i) every edge $i$ in the full graph is associated with probability $p_i$ for the occurrence of a bit-flip error event denoted by this edge;
	(ii) probabilities of error events in different qubits or in different time steps are independent.
	
	Any subset of edges in the final matching can be interpreted as a corresponding pattern of error events. 
	By construction of the graph, different edges correspond to different single-qubit errors. 
	Thus, taking into account the independence of error events, we can simply multiply probabilities of edges to get the probability of the error pattern denoted by this set of edges. 
	We infer that if an event is denoted by a subset of edges with indices $\{i_k\}_1^K$, then the probability of such event is $P(\{i_k\}_1^K) = \Pi_{k = 1}^{K}p_{i_k}$, where $p_{i_k}$ is the probability associated with the edge $i_k$.
	
	We want to show that maximization of probability of an error pattern denoted by a subset of edges is equivalent to minimization of the sum of weights of these edges. 
	Weight $w_i$ of the edge $i$ is defined as the following function of the probability $p_i$:
	
	\begin{equation}
		w_i = -\log p_i.
	\end{equation}
	
	The desired result is a simple consequence of the following proposition.
	
	\textbf{Proposition.} 
	Take any two subsets of edges. 
	Let us denote the first subset by edge's indices $\{i_k\}_1^{K_1}$ and the second subset by indices $\{j_k\}_1^{K_2}$. 
	The following statements are equivalent:
	\begin{enumerate}
		\item The probability of the event denoted by the subset of edges $\{i_k\}_1^{K_1}$ is greater than or equal to the probability of the event $\{j_k\}_1^{K_2}$:
		\begin{equation}	
			P(\{i_k\}_1^{K_1})\geq P(\{j_k\}_1^{K_2}).
		\end{equation}
		\item The sum of weights of the subset $\{i_k\}_1^{K_1}$ is less than or equal to the sum of weights of the subset $\{j_k\}_1^{K_2}$:
		\begin{equation}	
			\sum_{i_k} w_{i_k} \leq \sum_{j_k} w_{j_k}.
		\end{equation}
	\end{enumerate}
	
	\textbf{Proof.} Let us consider the following set of relations:
	\begin{equation}
		\sum_{i_k} w_{i_k}\leq\sum_{j_k} w_{j_k}\Leftrightarrow
	\end{equation}
	\begin{equation}
		\Leftrightarrow-\sum_{i_k}\log p_{i_k}\leq-\sum_{j_k}\log p_{j_k}\Leftrightarrow
	\end{equation}
	\begin{equation}
		\Leftrightarrow\log \Pi_{i_k}p_{i_k}\geq\log \Pi_{j_k}p_{j_k}\Leftrightarrow
	\end{equation}
	\begin{equation}
		\Leftrightarrow\log P(\{i_k\}_1^{K_1})\geq\log P(\{j_k\}_1^{K_2})\Leftrightarrow
	\end{equation}
	\begin{equation}
		\Leftrightarrow P(\{i_k\}_1^{K_1})\geq P(\{j_k\}_1^{K_2}).
	\end{equation}

	\section{Five-qubit repetition code with single ancilla and circular connectivity: realization and decoding}\label{sec:five-decoding}
	
\begin{center}
		\begin{figure}[!ht]
			\includegraphics[scale=0.8]{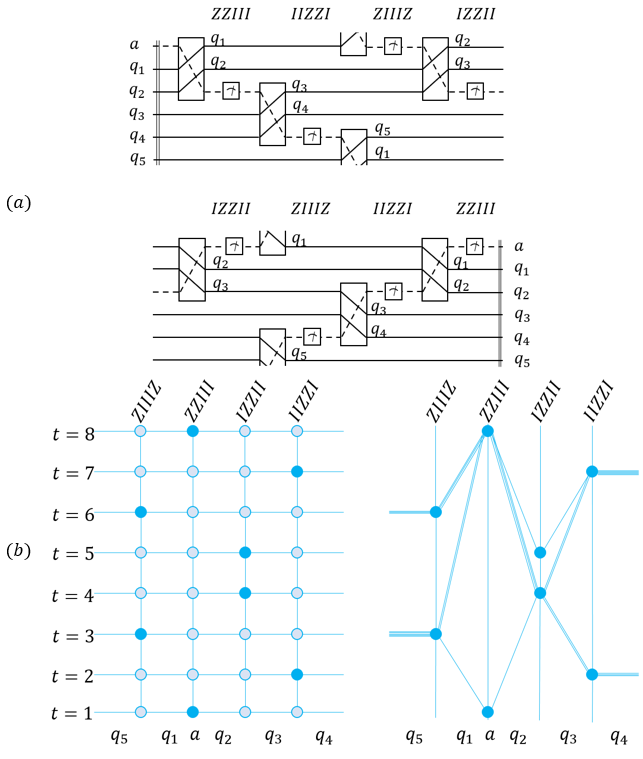}
			\caption{In (a) the circuit implementing the five-qubit repetition code is shown. 
				In (b) the implementation of the decoding scheme for the five-qubit repetition code is demonstrated.}
			\label{RepetitonGeneralizedCircuitGraph}
		\end{figure}
	\end{center}

	Here we consider repetition code with 5 data qubits and the {generating set} $\{ZZIII,IZZII,IIZZI,ZIIIZ\}$. 
	This particular form of the {generating set} allows to use the circuit illustrated in Fig.~\ref{RepetitonGeneralizedCircuitGraph}(a) with single ancilla, linear connectivity, and two-qubit gates of the CNS type only. 
	Decoding graph in this case is shown in Fig.~\ref{RepetitonGeneralizedCircuitGraph}(b).
	
	\begin{center}
		\begin{figure*}[!ht]
			\includegraphics[scale=0.7]{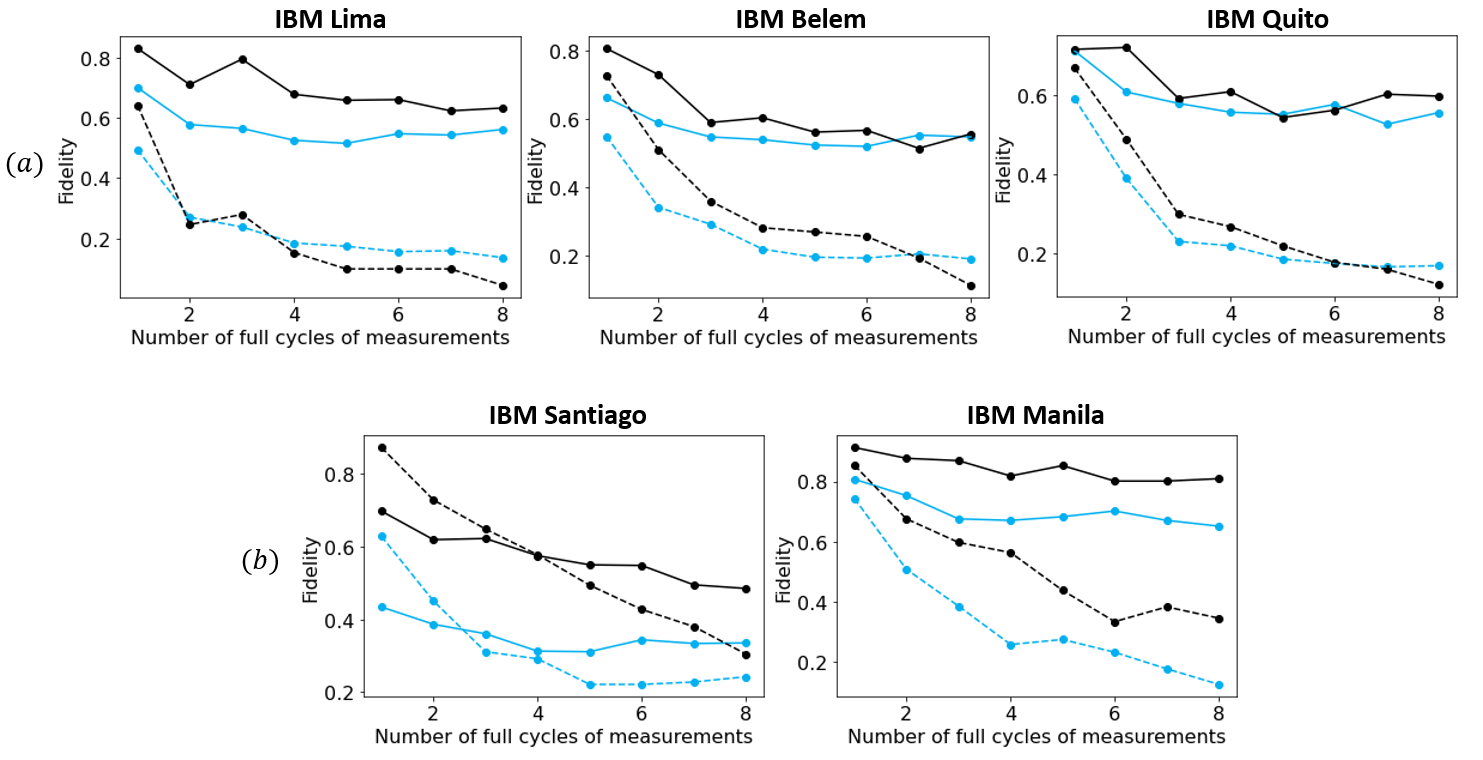}
			\caption{Comparison of fidelities in experiments on quantum processors.
				Our scheme of repetition code using 4 physical qubits is compared with the well-known benchmark scheme of repetition code using 5 qubits. 
				Each point on the graph corresponds to an estimated fidelity of a state in 1 experiment with a given number of full cycles of measurements. 
				Blue (light gray) lines represent results of experiments with the realization of our scheme and black lines represent results for the benchmark realization of the same repetition code. In our scheme, 4 physical qubits (3 data qubits and 1 ancilla qubit) are used, and in the benchmark scheme, 5 physical qubits (3 data qubits and 2 ancilla qubits) are used.
				Dotted lines depict fidelities of resulting states without correction. Experiments in (a) correspond to quantum processors with Y-type connectivity; experiments in (b) correspond to quantum processors with I-type connectivity.}
			\label{IBM_Experiments}
		\end{figure*}
	\end{center}
	
	\begin{center}
		\begin{figure}[!ht]
			\includegraphics[scale=0.35]{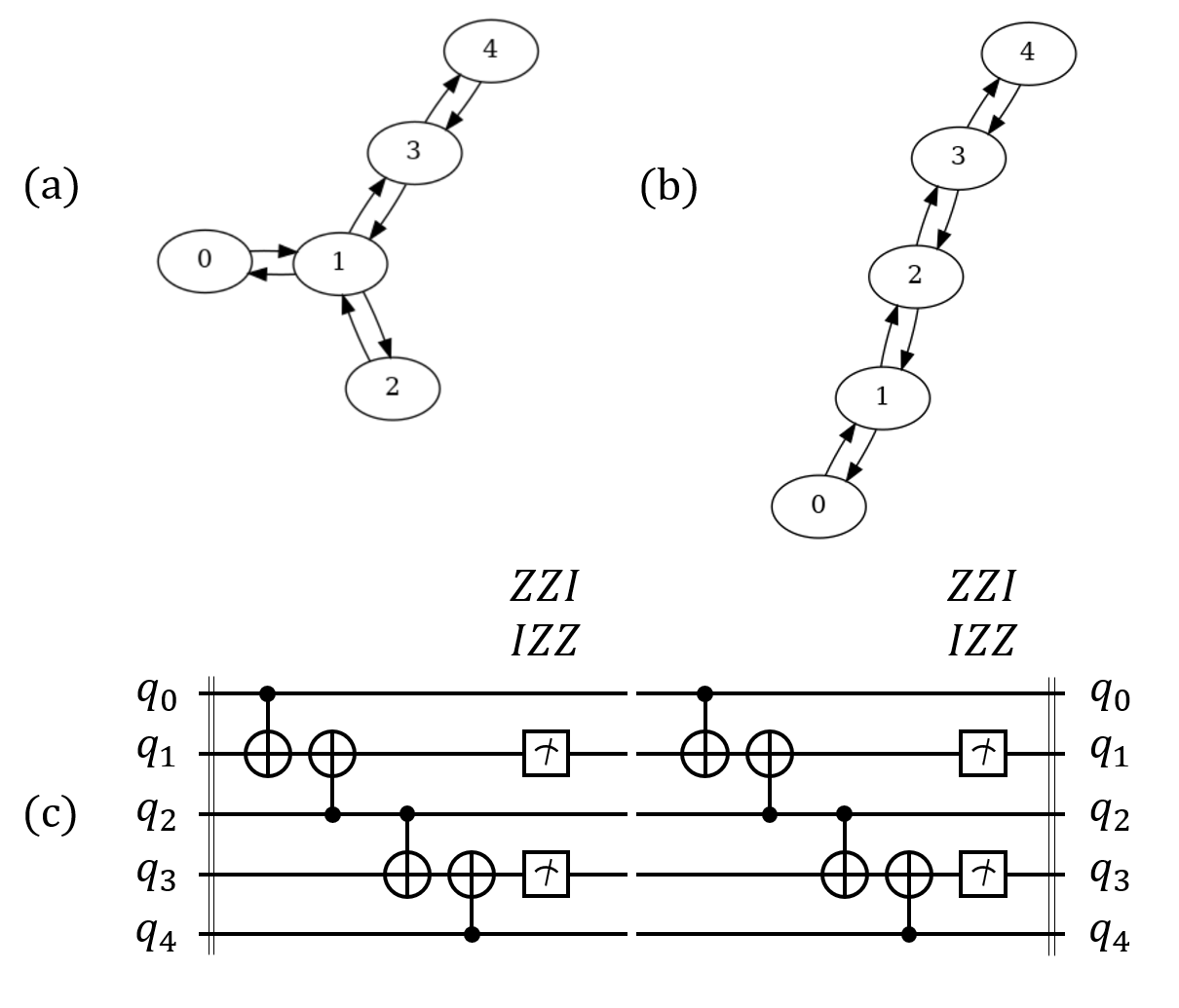}
			\caption{In (a) and (b) we illustrate the enumeration of physical qubits in connectivity patterns of the investigated quantum processors. 
				In (c) the circuit realizing 1 cycle of consecutive syndrome measurements for the 3-qubit repetition code with 3 data qubits and 2 ancilla qubits is presented.}
			\label{5QRepetitionScheme}
		\end{figure}
	\end{center}

	\section{Repetition code with two ancillas}\label{sec:five-qubit-repetition-two-ancillas}
	
	The well-known scheme realizing single cycle of syndrome measurements for the repetition code with 3 data qubits and 2 ancilla qubits is presented in Fig.~\ref{5QRepetitionScheme}. 
	One cycle contains four generators' measurements to match one cycle of measurements in our scheme (Fig.~\ref{Schemes_3q}(c)).
	
	\section{Experiments: single ancilla circular connectivity scheme and the repetition code with two ancillas}\label{sec:two-ancillas-experiments}
	
	We analyze performance of our scheme using 4 physical qubits (3 data qubits and 1 ancilla qubit) and compare it with the well-known repetition code scheme using 5 physical qubits (3 data qubits and 2 ancilla qubits) in Fig.~\ref{IBM_Experiments}. 
	Appendix~\ref{sec:five-qubit-repetition-two-ancillas} contains a description of the 5-qubit scheme. 
	IBM Bogota was not available at the moment of conducting the experiment. 
	Results of experiments show significant improvement in the fidelity of the preserved state on 4 out of 5 quantum processors, as is shown in Fig.~\ref{IBM_Experiments}. 
	
	We see that fidelities of corrected states for 4 physical qubits in our scheme are quite close to fidelities for 5 physical qubits in the well-known scheme, although lacking physical connection and realization of CNOT$\cdot$SWAP without CNS might have worsened potential performance of our scheme. Additionally, 5-qubit scheme is shorter in a sense that it contains less quantum operations per 1 cycle. 
	This means that the period of time for an execution of 1 cycle of measurements could be significantly lower in the case of 5-qubit scheme than in the case of 4-qubit scheme. 
	This, in turn, could lead to lower intensity of errors in the case of 5-qubit scheme. 
	So the fact that our 4-qubit scheme is still quite close in the performance to the 5-qubit scheme despite all the problems is a very positive result. 
	
	\section{Sequential scheme for the surface code with single ancilla and near-neighbour connectivity}\label{sec:surfacecode_1_22_steps}
	
\begin{center}
		\begin{figure*}[!ht]
			\includegraphics[scale=0.7]{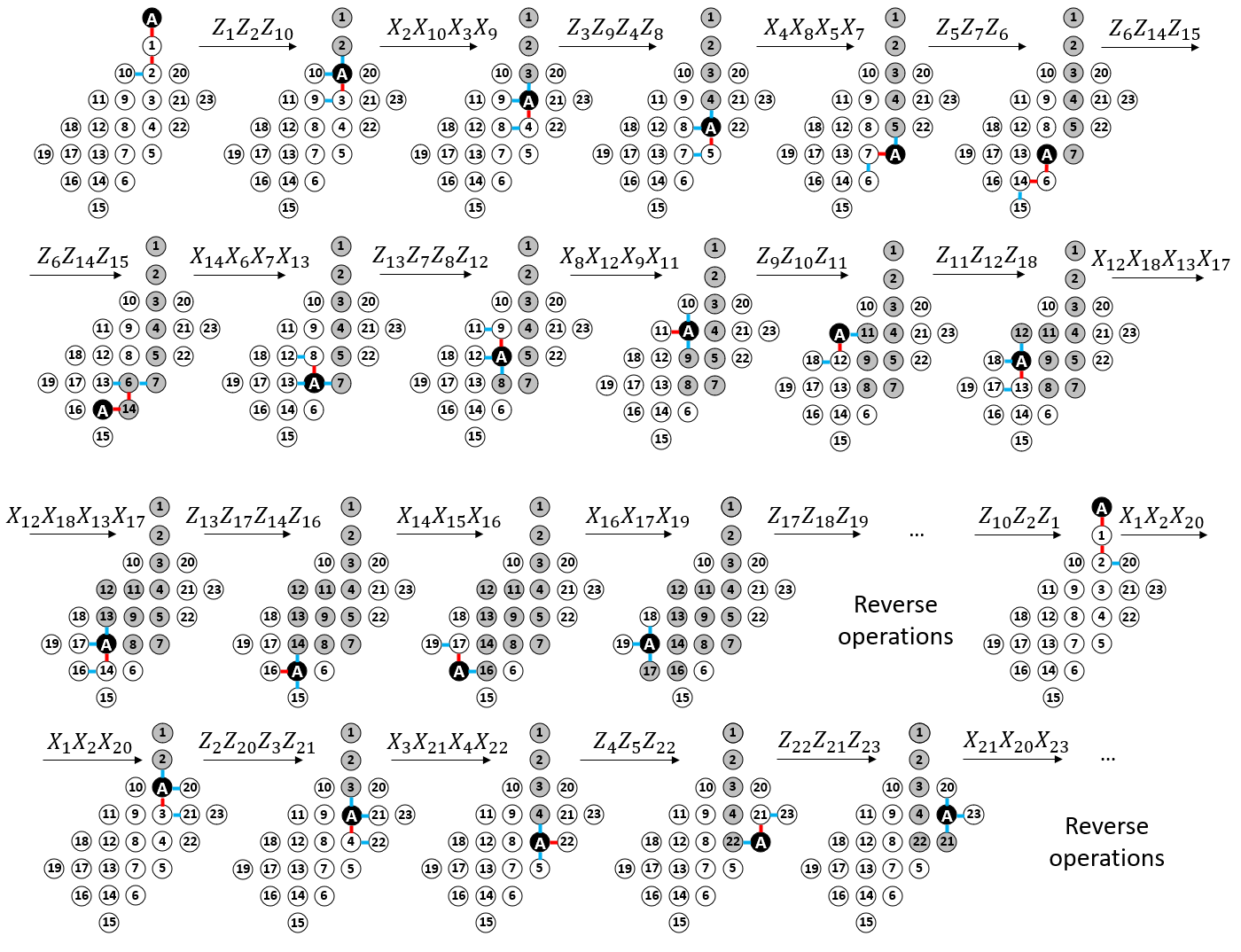}
			\caption{{Scheme for sequential measurements of Surface code's stabilizer with single ancilla. Red (dark gray) edge corresponds to CNS gate, and blue (light gray) edge corresponds to CNOT gate.}}
			\label{SurfaceCode_1_23}
		\end{figure*}
	\end{center}

	{Fig.~\ref{SurfaceCode_1_23} contains all the steps for the sequential process of measurements in the surface code layout from Fig.~\ref{SurfaceCode}. The first three steps and detailed circuits are given in Fig.~\ref{SurfaceCode_1_3} of the main text.}
	
	\section{Quantum processor characteristics}\label{sec:characteristics}
	
	Below are provided single-qubit noise characteristics of the qubit with index 0 in the IBM Belem quantum processor. 
	These characteristics are used to define noise channel in simulations:
	\begin{itemize}
		\item probability of single-qubit error for 1-qubit depolarizing channel $p_{D_1} = 0.000276$;
		\item time $T_1 = 78.11$ microseconds;
		\item time $T_2 = 114.09$ microseconds;
		\item single-qubit gate time $T_g = 35.55$ nanoseconds;
		\item probability of the measurement error $p_{meas} = 0.02$.
	\end{itemize}
	
	\bibliography{references.bib}
	
\end{document}